\documentclass[12pt,preprint]{aastex}

\newcommand{\msun}{M$_\odot$}

\newcommand{\spitzer}{\emph{Spitzer}}
\newcommand{\rspup}{RS~Pup}

\newcommand{\deltacep}{$\delta$~Cep}

\slugcomment{Accepted by ApJ on Nov 11, 2009}

\shorttitle{Spitzer observations of Galactic Cepheids}
\shortauthors{Marengo et al.}

\begin{document}


\title{Galactic Cepheids with Spitzer: I. Leavitt Law and Colors}


\author{M. Marengo\altaffilmark{1,2}, N. R. Evans\altaffilmark{2},
  P. Barmby\altaffilmark{3,2}, G. Bono\altaffilmark{4,5},
  D. L. Welch\altaffilmark{6} and M. Romaniello\altaffilmark{7}} 
\altaffiltext{1}{Dept. of Physics and Astronomy, Iowa State
  University, Ames, IA 50011} 
\altaffiltext{2}{Harvard-Smithsonian Center for Astrophysics,
  60 Garden St., Cambridge, MA 02138}
\altaffiltext{3}{Dept. of Physics and Astronomy, University of Western
  Ontario, London, Ontario, N6A 3K7 Canada}
\altaffiltext{4}{Dept. of Physics, Universit\`a di Roma Tor Vergata,
  via della Ricerca Scientifica 1, 00133 Roma, Italy} 
\altaffiltext{5}{INAF--Osservatorio Astronomico di Roma, via Frascati
  33,  00040 Monte Porzio Catone, Italy}
\altaffiltext{6}{Dept. of Physics and Astronomy, McMaster University,
  Hamilton, Ontario, L8S 4M1, Canada} 
\altaffiltext{7}{European Southern Observatory,
  Karl-Schwarzschild-Str. 2,  85748 Garching bei Munchen, Germany}




\begin{abstract}

  Classical Cepheid variable stars have been important indicators of
  extragalactic distance and Galactic evolution for over a
  century. The \spitzer{} Space Telescope has opened the possibility
  of extending the study of Cepheids into the mid- and far-infrared, where
  interstellar extinction is reduced. We have obtained photometry from
  images of a sample of Galactic Cepheids with the IRAC and MIPS
  instruments on \spitzer. Here we present the first mid-infrared
  period--luminosity relations for Classical Cepheids in the Galaxy,
  and the first ever Cepheid period--luminosity relations at 24 and
  70~\micron. We compare these relations with theoretical predictions,
  and with period--luminosity relations obtained in recent studies of
  the Large Magellanic Cloud. We find a significant period--color
  relation for the $[3.6]-[8.0]$ IRAC color. Other mid-infrared colors for
  both Cepheids and non-variable supergiants are strongly affected by
  variable molecular spectral features, in particular deep CO
  absorption bands. We do not find strong evidence for mid-infrared excess
  caused by warm ($\sim 500$~K) circumstellar dust. We discuss the
  possibility that recent detections with near-infrared interferometers of
  circumstellar shells around $\delta$~Cep, $\ell$~Car, Polaris, Y~Oph
  and RS~Pup may be a signature of shocked gas emission in a dust-poor
  wind associated to pulsation-driven mass loss.

\end{abstract}



\section{Introduction}\label{sec-intro}

Although it was over 100 years ago that Henrietta Leavitt discovered
the Cepheid Period--Luminosity relation (PL, \citealt{leavitt1908}),
or ``Leavitt Law'', few tools in astronomy have such enduring
importance. Classical Cepheid variable stars are fundamental
calibrators of the extragalactic distance scale and in addition their
observed properties are a benchmark for stellar evolution models of
intermediate mass stars. It is now widely accepted that distances
derived from SN Ia are still significantly influenced by the classical
Cepheid calibration: \citet{riess2005} found a change of 15\% in the
value of $H_0$ when only modern, high-quality SN Ia data and HST
Cepheids were used. Recent works have extended optical and
near-infrared PL and period--color (PC) relations to the mid-infrared
(mid-IR), where there is less interstellar extinction, with
observations of Large Magellanic Cloud (LMC) Cepheids
(\citealt{freedman2008, ngeow2008, ngeow2009, madore2009a}) with the
\spitzer{} Space Telescope \citep{werner2004}.

Evolutionary and pulsation properties of intermediate-mass stars in
the core He-burning phase, like Cepheids, play a crucial role in
several long-standing astrophysical problems. They are transition
objects between stellar structures ending up their evolution either as
white dwarfs or as core collapse supernovae. Therefore, they are not
only the most popular primary distance indicators, but are also
crucial to understanding the chemical evolution of stellar systems
hosting a substantial fraction of young stars, e.g the Galactic disk
and the dwarf irregular galaxies in the Local Group and in the Local
Volume ($d \la 10$~Mpc).

Although these objects are fundamental for stellar evolution
\citep{bono2000, beaulieu2001}, stellar pulsation \citep{bono1999,
  marconi2005}, and Galactic chemical evolution models
\citep{pedicelli2009, spitoni2009}, current predictions are still
hampered by several problems. The most outstanding issue is the
``Cepheid mass discrepancy'' between pulsation masses of classical
Cepheids and their evolutionary masses. Evidence was brought forward
more than 30 years ago by \citet{fricke1972} who found that pulsation
masses were from 1.5 to 2 times smaller than the evolutionary
masses. This conundrum was partially solved \citep{moskalik1992} by
the new sets of radiative opacities released by the Opacity Project
\citep{seaton1994} and by OPAL \citep{rogers1992}. However, several
recent investigations focused on Galactic Cepheids \citep{bono2001a,
  caputo2005} and Magellanic Cloud Cepheids \citep{beaulieu2001,
  bono2002, keller2006} suggest that such a discrepancy still amounts
to 10--15\%. Measured masses of Galactic binary Cepheids
(e.g. \citealt{evans2008}) are also smaller than predicted by
evolutionary models neglecting core convective overshooting during
central hydrogen burning phases.

The relative importance of the main factors affecting Cepheid model
mass estimates (extra-mixing, rotation, radiative opacity, mass loss
and binarity) is still debated. Even though commonly used
semi-empirical relations \citep{reimers1975, dejager1997} do not
predict enough mass loss to solve the Cepheid mass discrepancy
problem, mass loss may indeed be the key culprit among the physical
mechanisms suggested to explain the mass discrepancy problem. The
semi-empirical mass loss relation derived by \citet{reimers1975} is
clearly inadequate to correctly estimate the mass loss for certain
evolutionary phases of giant stars (see e.g. \citealt{willson2000}). 
A plausible increase in the typically adopted Reimers wind free
parameter ($0.2 \le \eta \le 0.4$) does not account for the entire
range in mass covered by cluster horizontal branch (central helium
burning) stars \citep{yong2000, castellani2005, serenelli2005}. 
There is no good reason why a similar discrepancy should not 
affect intermediate-mass stars.

Mass loss is often betrayed by the presence of dust shells,
detectable as infrared excesses, or by stellar winds which produce
blue-shifted absorption dips in the ultraviolet. Empirical estimates
of mass loss rates based on infrared (IRAS) and ultraviolet (IUE
spectra) observations for a large sample of Galactic Cepheids suggest
mass-loss rates ranging from 10$^{-10}$ to 10$^{-7}$~\msun~yr$^{-1}$
\citep{deasy1988}. However, evidence for mass loss rates high enough
to affect evolution is very rare, and it is not clear that mass loss
is a wide-spread phenomenon. \citet{mcalary1986} used IRAS photometry
\citep{beichman1985} and found evidence of very cool dust ($T_d \la
50$~K) around two classical Cepheids (\rspup{} and SU~Cas) known to be
associated with reflection nebul\ae{}. Due to the very large IRAS
beamsize (as large as $\sim 5$~arcmin), however, it was very difficult
to separate local dust emission from ``Galactic cirrus'' background
emission. More recently, $K$ band near-IR interferometric observations
\citep{merand2006, merand2007, kervella2006, kervella2008,
  kervella2009} have detected circumstellar emission around five
Classical Cepheids: \deltacep{}, $\ell$~Car, Polaris, Y~Oph and
RS~Pup. While the nature of the material responsible for this emission
remains mysterious, this is a tantalizing suggestion that mass loss
activity may be present around these nearby Cepheids. A similar
conclusion was also reached by \citet{neilson2009}, that using OGLE
(Optical Gravitational Lensing Experiment; \citealt{udalski1999}) and
\spitzer{} data for Magellanic Cepheids, found evidence of a wide
range of mass loss rates.

To investigate this possibility, we have obtained \spitzer{}
observations of a sample of Galactic Classical Cepheids. All stars
were observed with both the \spitzer{} Infrared Array Camera (IRAC,
\citealt{fazio2004}) and the Multiband Infrared Photometer for
\spitzer{} (MIPS, \citealt{rieke2004}). The aim of the IRAC
observations was to characterize Galactic Cepheid colors in the
mid-IR, where the emission from the photosphere is still dominant, and
search for infrared excess related to warm ($\sim 500$~K)
circumstellar dust. The MIPS observations were intended to investigate
the presence of extended emission from cool ($\la 100$~K) dust, taking
advantage of the higher angular resolution of \spitzer{} (5~arcsec at
24~\micron). In this paper we present the photometry of our Cepheid
sample in the IRAC and MIPS bands, discussing their PL and
period-color (PC) relations. The results of our search for extended
emission in IRAC and MIPS images will be presented in a separate paper
(Barmby et al. in preparation).

The criteria for our sample selection are laid out in
section~\ref{sec-sample}, and the observations are described in
section~\ref{sec-obs}. The techniques adopted to measure the source
photometry in all IRAC and MIPS bands are discussed in detail in
section~\ref{sec-phot}. In section~\ref{sec-leavitt} we derive the
Leavitt Law and PC relations in the IRAC and MIPS bands, and we
compare them with similar relations obtained for the LMC. We study the
intrinsic mid-IR colors of Cepheids in section~\ref{sec-excess}, where
we set limits on infrared excess at \spitzer{} wavelengths. Our
results are discussed in section~\ref{sec-discussion} and summarized
in section~\ref{sec-summary}.

\section{Sample Selection}\label{sec-sample}

Our sample stars are listed in Table~\ref{tab-targets}.  They include
29 bright $(K<5.1)$ nearby Cepheids, as well as 3 non-variable
supergiants and one red giant added for comparison.  The sample was
selected to cover a range of several characteristics which may
influence mass loss, or enable us to understand it better. Several of
the Cepheids are in clusters, three are first-overtone pulsators, and
eight are in multiple star systems. The range of periods is about a
factor of ten, and several Bump Cepheids ($9 \le P \le 12$ days) are
included so that we could investigate whether these objects present a
peculiar mass loss rate.  Our sample includes all 9 Cepheids with
measured angular diameters \citep{nordgren2000, lane2002,
  kervella2004} at the time of the \spitzer{} proposal submission, as
well as 5 stars (Polaris, S~Mus, W~Sgr, V350~Sgr, and FF~Aql) with
observed masses \citep{evans1997, evans2006, evans2008, evans2009,
  benedict2007}. V350~Sgr, S~Mus, RS~Pup, and $\beta$~Dor are the only
Cepheids with 60~\micron{} excesses observed by IRAS, so they can be
used to check the earlier results. Our sample also includes Y~Oph, a
long period Cepheid with a surprisingly small amplitude and placid
velocity structure (based on line profiles). Its unusual
characteristics make it a candidate for an unusual history (either
mass loss, coalescence, peculiar chemical composition, evolutionary
status).

The stars were also selected for the availability of accurate distance
determinations (with the exception of the cluster star V636~Sco whose
parallax is very uncertain). For 7 Cepheids we have adopted the HST
Fine Guidance Sensor parallaxes measured by \citet{benedict2007},
while we have adopted the Hipparcos distance for Polaris
\citep{vanleeuwen2007}. For the other stars we have adopted distances
derived with the infrared surface brightness (hereafter IRSB)
technique (see e.g. \citealt{fouque1997}). While the IRSB method can
provide very precise parallaxes when the pulsational velocity and
lightcurve of a pulsating star is known, the conversion between the
observable radial velocity $v_r$ into pulsational velocity $v_p$
depends on the so-called projection factor $p = v_p/v_r$. The
$p$-factor, despite its name, does not depend exclusively on the
geometry of the star, but also on its physical properties, including
the period (see e.g. \citealt{nardetto2007}). Several $p(P)$ relations
have been proposed, and agreement on which one better describes the
true dependence of the $p$-factor from period is missing (see
e.g. \citealt{romaniello2008}). Until 2007 several catalogs of Cepheid
IRSB distances (\citealt{storm2004, groenewegen2004, barnes2005} among
others) used the relation $p = 1.39 - 0.03 \log P$ \citep{gieren1993}
characterized by a weak dependence from the period. More recently
\citet{nardetto2007} proposed a new ``steeper'' relation $p = 1.366 -
0.075 \log P$, which is used in the \citet{fouque2007} catalog.

In Table~\ref{tab-targets} we provide two separate columns for the
distance of our targets: the first column mainly lists the ``new''
distance determination from \citet{fouque2007}, while the second lists
the ``old'' IRSB distances. In both cases, whenever available, we
adopt the \citet{benedict2007} and \citet{vanleeuwen2007} distances,
which are independent from the $p$-factor. For three stars (DT~Cyg,
V350~Sgr and U~Aql) we could not find their IRSB distance estimated
using the ``old'' $p(P)$ relation. For completeness we derive the
``old'' distance of these stars (plus V636~Sco) using the method
described in \citet{fouque1997}. These distances, and the parameters
used to derive them, are listed in Table~\ref{tab-extrad}. $V$ and
$E(B-V)$ are derived from the Dunlap Observatory Database of Galactic
Classical Cepheids \citep{fernie1995}, $K$ magnitudes from
\citet{vanleeuwen2007} or \citet{welch1984}, while $R_0$ and $\Delta
R_0$ from \citet{moskalik2005}.

All stars except 3 (DT~Cyg, SZ~Tau and Polaris) are listed as
fundamental mode pulsators, according to \citet{fouque2007}. FF Aql is
here assumed to pulsate in the fundamental mode, as proposed by
\citet{benedict2007}, despite having been previously classified as a
first overtone Cepheid by \citet{feast1997}.

\section{Observations}\label{sec-obs}

The observations were executed between July 19, 2006 and October 28,
2007 as part of the \spitzer{} Space Telescope Cycle 3 General
Observer program with PID 30666. Each star was observed in IRAC 3.6,
4.5, 5.8 and 8.0~\micron{} and MIPS 24 and 70~\micron{} bands. The
MIPS 160~\micron{} band was not used: the known near-infrared leak in
this filter would have resulted in strong contamination by 
starlight.

The IRAC observations were designed to provide, whenever possible,
high S/N unsaturated images at all wavelengths well within the
linearity regime of the detectors, using the IRAC subarray mode with
0.02 or 0.4~sec frame times. For targets brighter than $K \sim 2.5$
even the shortest subarray frame time was too long to prevent
saturation. For this reason, 6 stars were observed using IRAC full
frame stellar mode (frame time of 0.4~sec at 3.6 and 4.5~\micron{} and
2~sec at 5.8 and 8.0~\micron), with the intent to produce heavily
saturated images suitable for PSF-fitting photometry on high S/N
unsaturated ``wings'' and diffraction spikes. Four other targets were
observed in both subarray and full frame stellar mode, to provide
photometric cross-calibration between the two modes of observations
and to check for linearity. Polaris was not observed with IRAC as part
of this program, since archival data obtained during Cycle 1
(\spitzer{} program 19) was available. Each star was dithered on the
array in 4 (subarray) or 5 (full frame) positions, using a small scale
Gaussian offsets pattern, in order to facilitate transient outlier and
bad pixel removals, and produce a better spatial sampling of the
images. The total time on source, for each band, was 25.6~sec for the
subarray observations. Full frame observations had 2.0~sec (3.6 and
4.5~\micron) and 10~sec (5.8 and 8.0~\micron) total integration
times. The IRAC data were reduced starting from the Basic Calibrated
Data (BCDs) generated by the \spitzer{} IRAC pipeline versions S14.4.0
through S16.1.0. Mosaic images with a pixel scale of 0.8627~arcsec/pix
for each source in each band were then created using the IRACproc
post-BCD software \citep{schuster2006}.

MIPS observations at both 24~\micron{} and 70~\micron{} were obtained
with the Photometry Astronomical Observation Template.  The
24~\micron{} observations used 1~cycle of 3~sec frames with the
standard 16 point dither pattern. The MIPS 70~\micron{} observations
used the default pixel scale and small field size, with 1 to 12 cycles
of 3 or 10~sec frames (depending on source and background brightness)
with the standard dither pattern. The cluster Cepheids S~Nor, U~Sgr,
and V~Cen had slightly different 70~\micron{} observations: 2 adjacent
fields of view were observed, to increase the sky coverage to $5\times
5$\arcmin.  The total on-source integration times were 48.2~sec at
24~\micron{} and from 37.7 to 1,300.2~sec at 70~\micron.  The MIPS
data, generated by the \spitzer{} pipeline version S16.1, was
retrieved from the Archive.  No post-processing was performed on most
of the data: post-BCD mosaics were used for the 24~\micron{} data and
most of the 70~\micron{} data (filtered versions were used).  For the
three cluster stars (S~Nor, SZ~Tau, U~Sgr), the 70~\micron{} data was
remosaiced using MOPEX to combine the data for the two fields of
view. The pipeline mosaics for two other stars (RS~Pup, GH Lup)
showed negative sidelobes; for these stars the BCDs were time- and
column filtered and the mosaics re-made.

\section{Source Photometry}\label{sec-phot}

\begin{figure}[h!]
 \centering
 \includegraphics[width=0.85\textwidth, angle=-90]{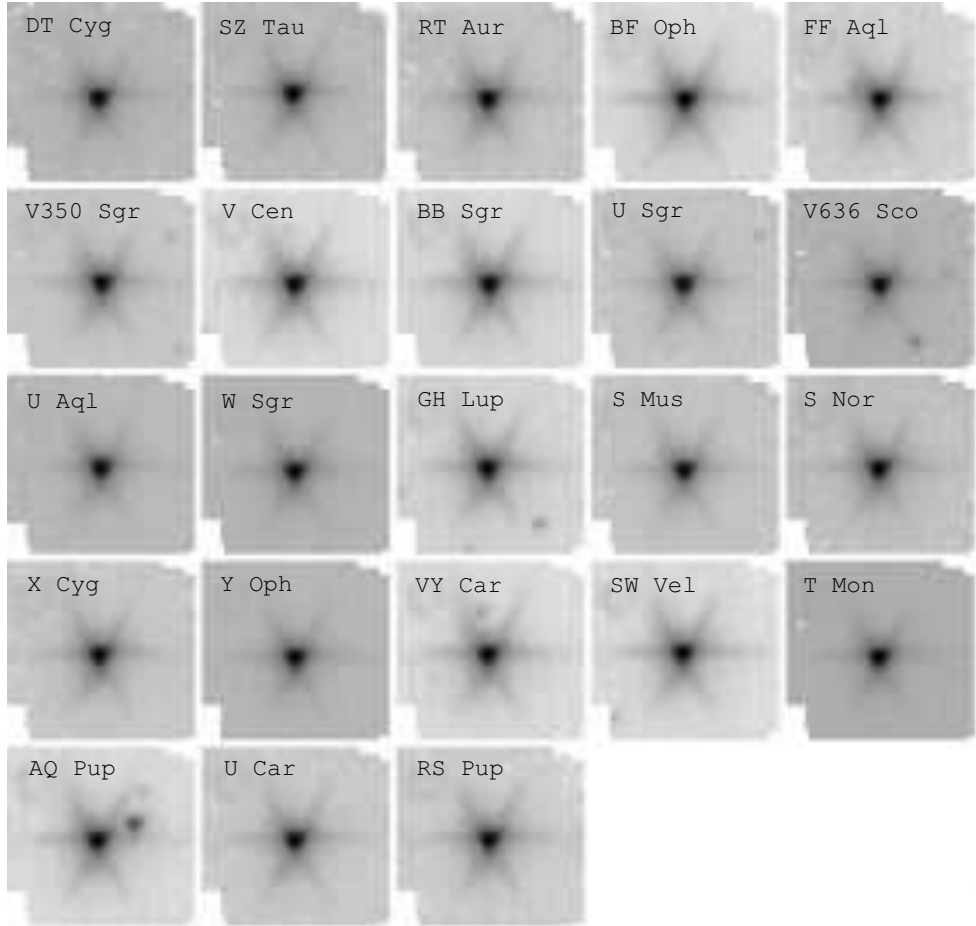}
 \caption{Thumbnail images of the 23 Cepheids observed in the IRAC
   subarray mode at 3.6~\micron, shown with a logarithmic color
   scale. The size of each thumbnail is about $45 \times
   45$~arcsec. Only three targets (FF~Aql, VY~Car and AQ~Pup) show one
   or more background stars within the standard 12.2 arcsec radius
   aperture used to measure the stars photometry. For these three
   stars a smaller 6.1~arcsec radius aperture has been
   used.}\label{fig-thumb}
\end{figure}

We measured the IRAC Vega magnitude of all sources observed in
subarray mode using aperture photometry. We derived photometric zero
points from the IRAC absolute calibration factors (FLUXCONV) in the
BCD headers (listed in Table~\ref{tab-params}). For most sources we
used an aperture of 12.2~arcsec radius ($\sim 10$~pixels in the IRAC
array pixel scale) with background level determined from a sky annulus
with 12.2 and 24.4 arcsec inner and outer radii. This aperture is the
standard used for the IRAC point source calibration, and as such
corresponds to an aperture correction equal to one. Three stars
(AQ~Pup, VY~Car and FF~Aql, see Figure~\ref{fig-thumb}) had one or
more background sources falling within this aperture, and required
using a smaller 6.1~arcsec radius aperture ($\sim 5$ IRAC pixels). For
these sources we derived our own aperture corrections, listed in
Table~\ref{tab-params}, using the other stars as templates. This
determination of the aperture correction is more reliable than using
the values listed in the IRAC Data
Handbook\footnote{http://ssc.spitzer.caltech.edu/irac/dh/} (2006),
which were obtained from images with different spatial sampling and
dither pattern. We have corrected our subarray point source photometry
for the position-dependent geometric and gain distortion factors
provided by \citet{hora2008}. These factors ($f_{corr}$) are listed in
Table~\ref{tab-params} for the subarray central position. The
photometric uncertainty was estimated adding in quadrature the sky RMS
variation and source photon noise. At 3.6 and 4.5~\micron, an extra
1\% uncertainty due to uncharacterized pixel-phase variations was also
added (see the IRAC Data Handbook, 2006, p. 50).

The IRAC photometry of the stars observed in full frame stellar mode
was obtained by fitting the unsaturated ``wings'' and diffraction
spikes of the mosaic images with the IRAC extended Pixel Response
Function (PRF). The extended PRF was constructed from multi-epoch deep
images of Sirius, Vega (2 epochs), Fomalhaut (2 epochs),
$\epsilon$~Eridani (2 epochs) and $\epsilon$~Indi, obtained in a
separate Guaranteed Time Observation program (PID 90) and IRAC
calibration observations. The individual observations were reduced
with IRACproc, and the mosaic of each star was masked to remove any
pixel closer than 50\% or 80\% to the saturation level to preserve
flux linearity. The individual images were then rescaled to one of the
images of Vega and combined with a sigma-clipping algorithm in order
to remove background sources. The final product is a high S/N
representation of the extended features of the IRAC Point Spread
Function (PSF) projected on the IRAC pixel grid (e.g. a PRF), and is
ideally suited to derive PSF fitting photometry of bright saturated
sources. Given that the PRF is normalized as a calibrated image of
Vega, the best-fit scaling factor between a star image and the PRF is
equal to the flux ratio of the star with Vega, and can be directly
converted into Vega magnitudes without relying on the IRAC absolute
calibration. Magnitudes derived with this procedure are not affected
by the position-dependent geometric and gain distortion factors, since
the sources and the PSF stars have all been observed at the center of
the IRAC arrays. An image of the IRAC PRF obtained with a similar
procedure is shown in \citet{marengo2009a}, and the PRF files are
available at the Spitzer Science Center web
site\footnote{http://ssc.spitzer.caltech.edu/irac/psf.html}.

The photometric error was estimated by bracketing the best subtraction
with an over- and under-subtracted fit in which the PSF-subtraction
residuals were of the same magnitude as the background noise.  The fit
was done using only the overlapping region where both the objects in
our programs and the PRF were not saturated. This region is relatively
narrow (width of $\sim 20$ -- 30 pixels) at 3.6 and 4.5~\micron, due
to the short total exposure time (2~sec) in these two bands. As a
consequence, The PSF-fitting photometric errors tend to be larger at
3.6 and 4.5~\micron{} than at 5.8 and 8.0~\micron. The dominant source
of uncertainty for the PSF-fitting photometry is the ``sampling
noise'' (grid pattern due to the coarse sampling of the IRAC PSF on
the pixel grid) in the subtraction residuals. This coherent noise does
not follow Poisson statistics. For this reason the photometric errors
estimated for the PSF-fitting photometry should not be taken as $1
\sigma$ statistical errors (as in the case of the aperture photometry
uncertainties), but rather as upper and lower limits in the
photometric measurements.

\begin{figure}[h!]
 \centering
 \includegraphics[width=0.65\textwidth, angle=-90]{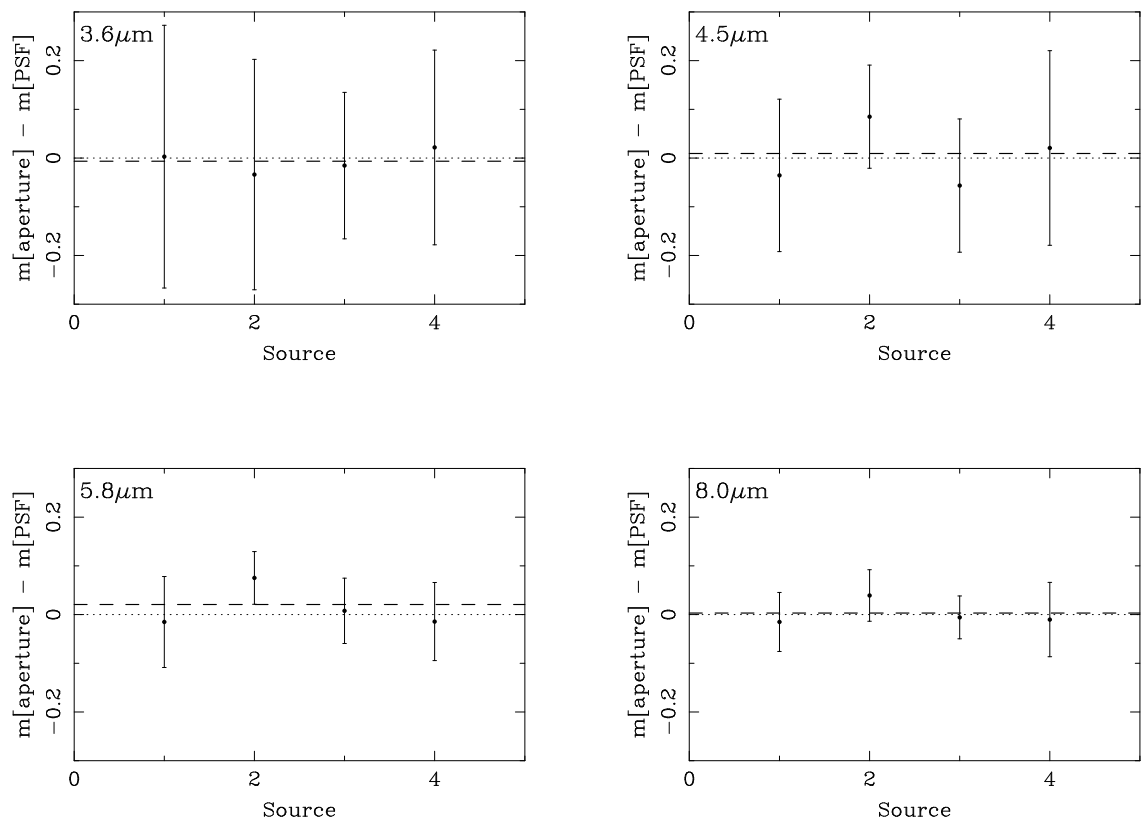}
 \caption{Comparison between aperture and PSF fitting photometry of
   four stars in our sample: \#1 V~Cen, \#2 U~Sgr, \#3 S~Nor and \#4
   SZ~Tau. The error bars are the combined uncertainty of the offset
   between the aperture and PSF-fitting magnitudes. The dashed line
   is the average offset between the two sets of measurements (much
   smaller than the photometric error of each data-point). Note that
   only V~Cen was observed in both subarray and full frame at the
   same epoch.}\label{fig-testpsf}
\end{figure}

To test the agreement between aperture and PSF-fitting photometry we
used the 4 stars observed in both modes. In general, the full frame
and subarray mode observations of these stars have been executed at
different epochs. As a consequence, based on Large Magellanic Cloud
(LMC) Cepheid photometry by \citet{madore2009b}, we should expect a
variability in the IRAC bands as large as $\pm 0.4$~mag. Still, we can
use these stars observed in both modes to assess the cross calibration
of the aperture and PSF-fitting methods, although with this
limitation. Table~\ref{tab-test} and Figure~\ref{fig-testpsf} show
that we do not measure in our data any statistically significant shift
between the subarray and the full frame photometry: the average
difference between subarray and full frame magnitudes is much smaller
than its RMS variation. The RMS variation (0.02, 0.07, 0.05 and
0.02~mags at 3.6, 4.5, 5.8 and 8.0~\micron{} respectively) is smaller
than the average uncertainty in the PSF-fitting photometry estimated
with the fit-bracketing method. For V~Cen (the only case in which both
subarray and full frame observations were executed at the same epoch),
the difference between the aperture and PSF fitting photometry is
0.01, $-$0.04, $-$0.02 and $-$0.02~mags at 3.6, 4.5, 5.8 and
8.0~\micron{} respectively. This is within the uncertainty of the PSF
fitting photometry of this star (0.27, 0.16, 0.09 and 0.06~mags,
respectively). At least in the case of V~Cen, this test can be viewed
as an independent confirmation of the IRAC absolute photometric
calibration, since aperture photometry depends on the FLUXCONV
parameters, while PSF-fitting with our PRF does not.

Table~\ref{tab-phot} shows the final magnitudes of all sources. When
available, we list the IRAC aperture photometry of unsaturated images
(which has smaller nominal uncertainty) rather than the PSF-fitting
photometry. The PSF-fitting photometry is instead listed for the
sources observed only in IRAC full frame mode.
  
Measurements of MIPS 24~\micron{} flux densities followed standard
procedures. Flux densities for most objects were measured in a
20\arcsec{} radius apertures with background subtraction from sky
annuli of 40\arcsec--50\arcsec{} and the recommended aperture
correction of 1.13 applied. There are a few stars with apparent
extended emission nearby; for these, flux densities were measured in
7\arcsec{} radius apertures with the same sky annulus, and an aperture
correction of 1.61. The flux densities measured for Polaris and
$\ell$~Car ($\sim3.7$~Jy) are formally above the saturation limit for
3~second exposures, but both objects are below the ``hard'' saturation
limit (4.1~Jy): the pipeline correctly replaced the saturated pixels
with values from the 0.5~second exposures taken as part of the
photometry AOR. Uncertainties for the 24~\micron{} flux densities were
computed using the standard IRAF/PHOT formula as:

\begin{equation}
\sigma^2 = (F/G + A\sigma^2+A^2\sigma^2/A^{\prime})^{1/2}
\end{equation}

\noindent
where $F$ is the flux in image units (MJy~sr$^{-1}$), $G$ is the
effective gain (the product of the detector gain, individual frame
exposure time, and number of frames, divided by the FLUXCONV factor
which converts from DN~s$^{-1}$ to MJy~sr$^{-1}$), $A$ is the aperture
area, $A^{\prime}$ is the background annulus area, and $\sigma$ is the
standard deviation of image counts in the background area. The values
listed in Table~\ref{tab-phot} do not include the 2\% uncertainty in
the absolute calibration \citep{engelbracht2007}, the dominant
systematic error for these bright stars.

Measurements of MIPS 70~\micron{} flux densities also followed
standard procedures. Flux densities were measured in 16\arcsec{} radius
apertures fixed on the 24~\micron{} position, with sky annuli of
18\arcsec--39\arcsec{}, and an aperture correction of 2.04. There were
no saturation issues in the 70~\micron{} data. Uncertainties in the
70~\micron{} flux densities were calculated using a similar equation
to that for MIPS-24, except that for 70~\micron{} data Poisson noise
is negligible. Following \citet{carpenter2008}, the uncertainties are
increased by extra multiplicative factors of $\eta_{\rm corr} = 2.5$
(noise correlation between pixels due to re-sampling during
mosaicing), and $\eta_{\rm sky} = 1.5 $ (excess sky noise due to the
data-taking procedure). The absolute calibration uncertainty at
70~\micron{} is 5\% \citep{gordon2007}, again not included in the data
table. A number of stars were not detected at 70~\micron{}: for these
sources $3\sigma$ upper limits are recorded.

Determining the effects of interstellar extinction on \spitzer{} IRAC
and MIPS photometry is tricky, because of the difficulty of estimating
the reddening in the Galaxy, compared to that for an external
galaxy. Measurements of the Galactic extinction curve \citep{lutz1996,
  indebetouw2005, flaherty2007, roman2007} when scaled to IRAC
wavelengths give $A_{5.8 \mu m}/A_B \approx 0.04$
\citep{rieke1985, freedman2008}. We should thus expect reddening at
mid-IR wavelengths for all our nearby Cepheids to be negligible, and
for this reason we do not correct  our data for reddening.

\section{Leavitt Law  and Period-Color Relations}\label{sec-leavitt} 

The advantages of deriving a PL relation (the Leavitt Law) in the
mid-IR (rather than in the optical or near-IR) are manifold:
luminosity variation amplitudes, metallicity effects and interstellar
extinction are all expected to be smaller at longer wavelengths. A well
calibrated PL relation in the IRAC and MIPS bands is of great utility
for the available Galactic and extragalactic stellar photometric
catalogs obtained in the \spitzer{} legacy projects.

The PC relation at optical wavelength provides important diagnostics
for basic stellar parameters such as effective temperature and
metallicity (see e.g. \citealt{sandage2004, sandage2009, tammann2003}
for a comparative study of the PC relations between the Galaxy, the
LMC and the SMC). Extending the study of PC relations into the mid-IR
offers the chance to study the effects of these parameters on
different spectral features, at wavelengths where the interstellar
reddening is less important. A firm understanding of the ``intrinsic''
color properties of classical Cepheids, and their dependence on
stellar parameters and period, is also essential to search for the
presence of ``extrinsic'' color excess that may be related to
circumstellar dust emission.

\subsection{PL Relations}\label{ssec-pl}

An IRAC PL relation has already been obtained for Cepheids in the LMC
\citep{freedman2008, ngeow2008, ngeow2009, madore2009a}, using data
from the SAGE project \citep{meixner2006}, and was used to derive the
distance modulus for two galaxies in the Local Group, NGC~6822
\citep{madore2009b} and IC~1613 \citep{freedman2009}. Using our
photometry in Table~\ref{tab-phot} we can derive the same relation for
a sample of Cepheids in our Galaxy. These are the same Cepheids that,
by virtue of their independent distance determination, are at the base
of the PL relation zero-point calibration.

We derived the best-fit PL relations for all IRAC and MIPS data with a
linear least square fit method, weighed on the uncertainty of each
photometric point. For IRAC subarray data we use as weight the $3
\sigma$ photometric uncertainty, while for all other measurements we
use the uncertainty quoted in Table~\ref{tab-phot}. In particular, for
IRAC full frame photometry we use the PSF-fit ``bracketing'' interval,
which is similar to a $3 \sigma$ uncertainty. For MIPS 70~\micron{}
data we exclude data points for which only an upper flux limit is
available. Table~\ref{tab-PL} lists the coefficients of the best-fit
relations, in the form:

\begin{equation}
M = \alpha \left [\log P - 1.0 \right] + \beta
\end{equation}

\begin{figure}[h!]
\centering
\includegraphics[width=0.70\textwidth, angle=0]{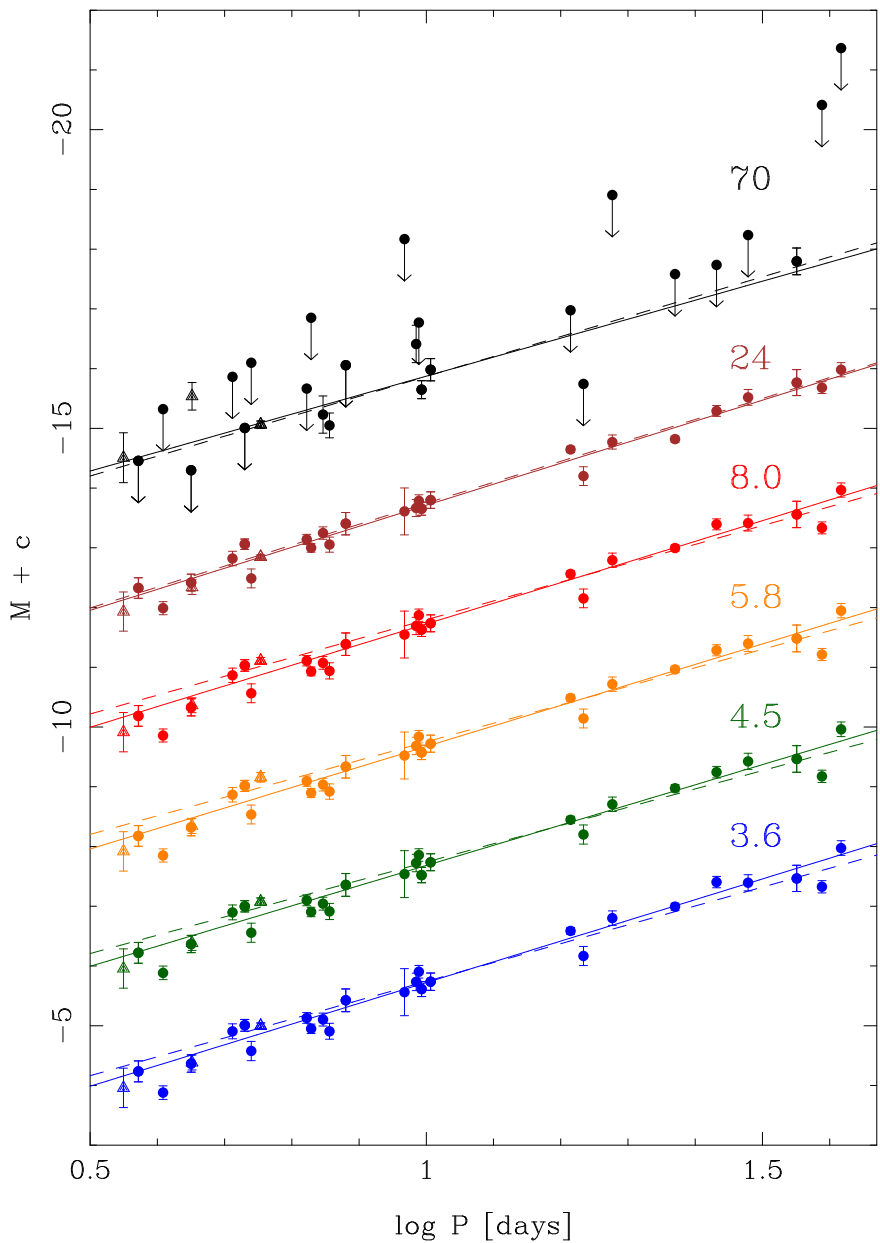}
\caption{Period vs. absolute magnitude plots for all IRAC and MIPS
  bands. Circles are fundamental mode Cepheids while triangles are
  first overtone (with period ``fundamentalized'' using the relation
  from \citealt{feast1997}). This plot shows $3 \sigma$ error bars
  for IRAC subarray photometry and the uncertainty listed in
  Table~\ref{tab-targets} for all other datapoints. The solid line
  shows the best-fit PL relation using all the distances from
  \citet{fouque2007}, while the dashed line uses only stars with
  astrometric distances.}\label{fig-PL}
\end{figure}

Note how the PL relations we have derived are based on single epoch
measurements, and not on mean photometry that would have required
multiple epoch observations. To include the first overtone Cepheids
(DT~Cyg, SZ~Tau and Polaris) in the fit, we have converted their
observed period $P_1$ into a ``fundamentalized'' period $P_0$ using
the relation:

\begin{equation}
P_1/P_0 = 0.716 - 0.027 \log P_1
\end{equation}

\noindent
from \citet{feast1997}, derived in turn from \citet{alcock1995}. We
provide the best-fit PL relations for three different cases. First,
using the ``new'' IRSB distance determination relying on the ``steep''
$p(P)$ relation. Then, using only the stars for which the
astrometric \citet{benedict2007} and \citet{vanleeuwen2007}
distances are available. Last, using the ``old'' distances based on the
$p(P)$ relation with weak period dependence. Note than in both ``old''
and ``new'' cases we still use the astrometric distances when
available. The best-fit PL relations for the ``new'' (solid lines) and
astrometric (dashed lines) distances are shown in
Figure~\ref{fig-PL}.

Table~\ref{tab-PL} also includes the $K$ band relations from
\citet{fouque2007}, \citet{benedict2007} and \citet{storm2004} for the
``new,'' astrometric, and ``old'' distances respectively. The
table shows that the slopes and zero points of the three
cases are consistent within their uncertainties. The astrometric
PL relations have larger uncertainties because most of the stars in
that sample have the less precise full frame PSF-fitting photometry,
and because of the smaller size of the sample. We can however note
some trends: the astrometric Leavitt laws are consistently 
less steep than the relations using IRSB distances. The ``old''
distances produce steeper relations than the ``new'' distances. A
similar trend is present in the $K$ band relations. The zero points
are very similar for all cases, well within their uncertainties.

Figure~\ref{fig-PL} also shows that FF~Aql (data-point at $\log P =
0.73$) closely matches the general trend. If instead, as suggested by
\citet{feast1997}, the source was pulsating in the first overtone, its
``fundamentalized'' PL ratio would be a significant outlier. This
result supports the view that FF~Aql is indeed behaving as a
fundamental mode pulsator as suggested by \citet{benedict2007}.

The dispersion around the best-fit PL relations is $\sim 0.2$~mag in
the four IRAC bands and MIPS 24~\micron{}, and $\sim 0.4$~mag at
70~\micron{}. This is comparable with the expected dispersion at
\spitzer{} wavelengths, once our photometric and distance uncertainty
is taken into account. The larger uncertainty in the best-fit
parameters at 70~\micron{} is due to the larger photometric errors and
the smaller number of sources for which a measured flux is available
at this wavelength.

\begin{figure}[h!]
\centering
\includegraphics[width=0.75\textwidth, angle=0]{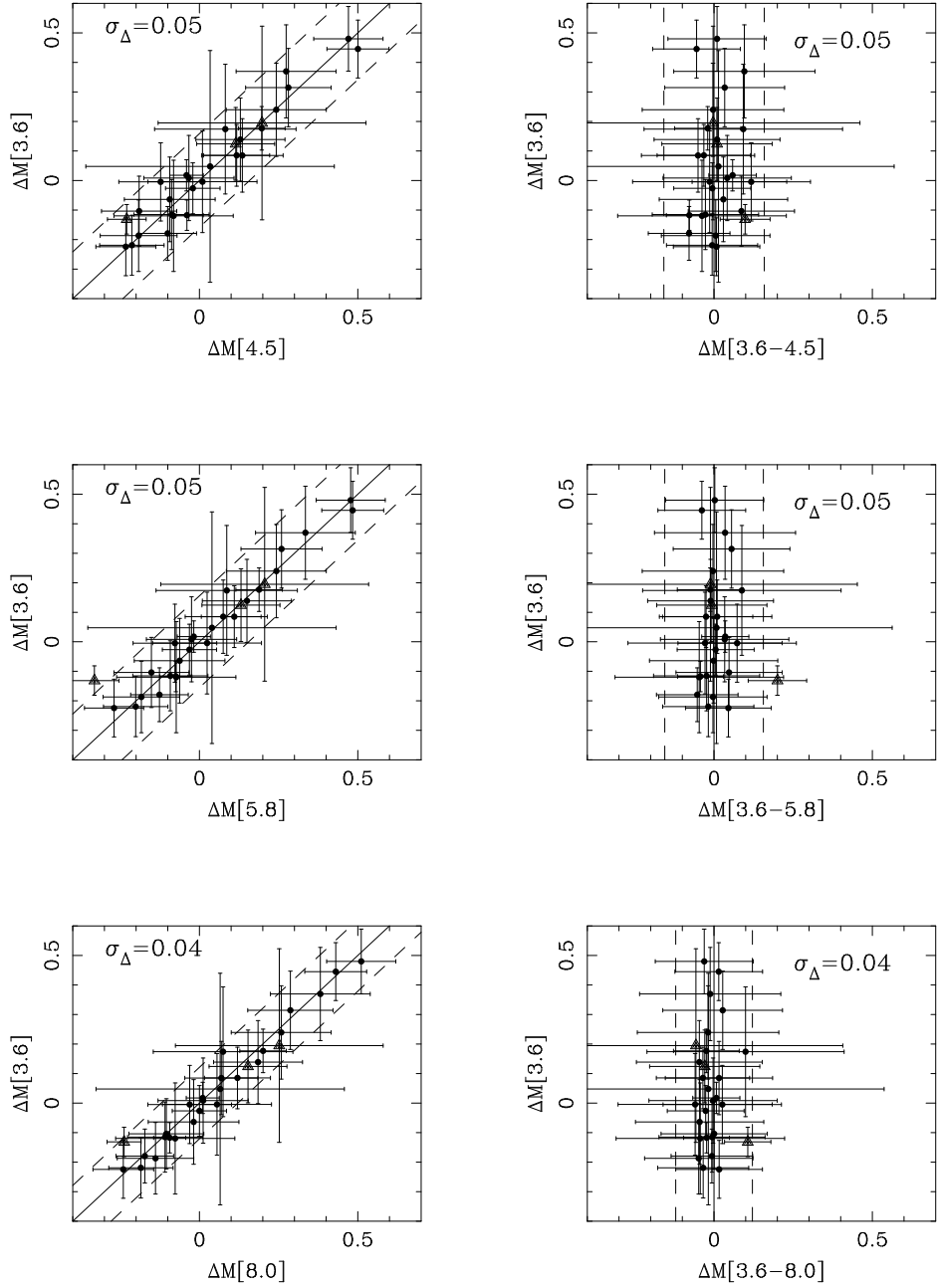}
\caption{\emph{Left panel} --- Correlated residuals from the PL
  relation. \emph{Right panel} --- Color residuals from the PL
  relation. Circles are fundamental mode Cepheids, while triangles are
  first overtone. The two dashed lines show the $3 \sigma_\Delta$
  scatter around the unity-slope line (solid line).}\label{fig-res}
\end{figure}

The dispersion around the best-fit PL relations has correlated and
uncorrelated components between bands. The correlated scatter is a
consequence of Cepheid variability. The uncorrelated scatter
$\sigma_\Delta$ is instead due in part to the absolute magnitude
uncertainties (geometric sum of the photometric error and the
uncertainty in the distance) and in part to intrinsic variations
related to the ``width'' of the Cepheid instability strip (due to
differences in metallicity and other stellar parameters).
Figure~\ref{fig-res} (left column) shows that the correlated component
of the fit residual has an amplitude of about $-0.3$ to $+0.5$~mag. We
may assume this value to approximate the amplitude of the Cepheid
light curves in the IRAC bands. While the weighted average residual is
zero (by virtue of the least-square fit used to derive the PL
relation), it is interesting to note that the IRAC residuals are
asymmetrically distributed. This may be a reflection of a prevalence
of asymmetric light curves in our sample. While the amplitude of the
correlated component of the PL dispersion in our data is similar to
the one measured in the LMC by \citet{freedman2008}, they do not note
any asymmetry in their sample (three times larger). Comparison between
our measured $\sigma_{\Delta}$ ($\sim 0.04$--0.05) and the absolute
magnitude uncertainty (on average $\sim 0.15$~mag) shows that we do not
have enough accuracy to measure the intrinsic scatter in the PL
relation, and all the observed scatter can be explained by the
uncertainty of our data.

Our sample does not show any correlation between variability and color
residuals in the IRAC bands (Figure~\ref{fig-res}, right column),
similar to the result obtained by \citet{madore2009a} in the
LMC. Again, we cannot test for color correlation between the IRAC and
MIPS bands due to the different epochs at which the two sets of data
were acquired.

\subsection{PC Relations}\label{ssec-PC}

\begin{figure}[h!]
\centering
\includegraphics[width=0.70\textwidth, angle=0]{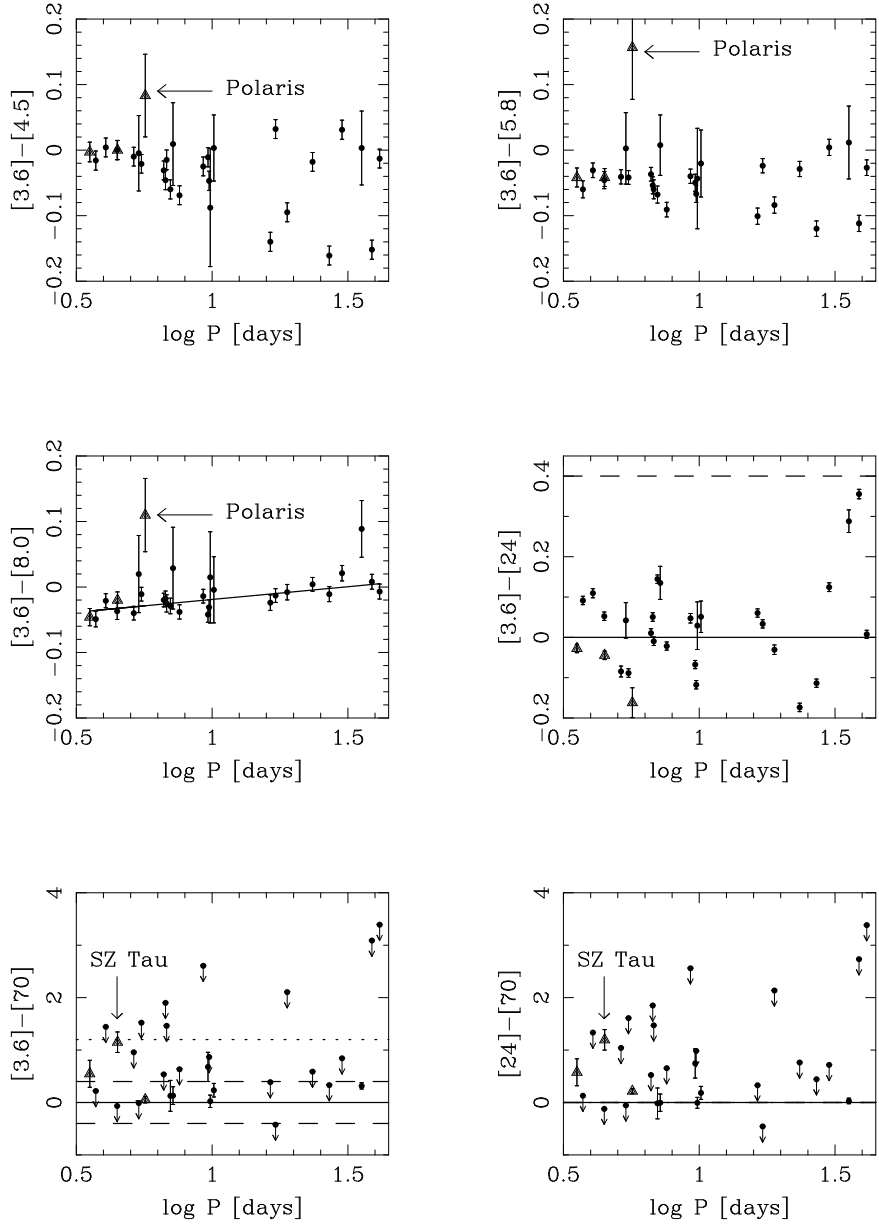}
\caption{\spitzer{} colors vs. period for our sample. The solid line
  in the $[3.6]-[4.5]$ color panel is the best fit PC relation. The
  solid line in the other panels, when present, is a marker for the
  zero color of Rayleigh-Jeans atmospheres). The dashed and dotted
  lines indicate the 1 and 3~$\sigma$ dispersion expected for the
  3.6~\micron{} lightcurve amplitude estimated in
  Section~\ref{ssec-pl}. Circles are fundamental mode Cepheids and
  triangles first overtone. SZ~Tau has a $[24]-[70]$ color excess
  above $3\sigma$ the noise level, and Polaris has consistently red
  ($\sim 0.1$~mag) IRAC colors.}\label{fig-pc}
\end{figure}

Figure~\ref{fig-pc} shows the period vs. \spitzer{} color trends in
our data. These diagrams are better suited to analyze color variations
between the sample Cepheids than the color residuals in
Figure~\ref{fig-res}, because they do not suffer from the larger
uncertainties due to the distance estimates necessary to derive the PL
fits. Even in the case of colors using only IRAC bands (taken
simultaneously, and hence with each star at a given phase), the data
show a scatter that is well above the photometric error. The only
exception is the $[3.6]-[8.0]$ color, for which it is possible to
derive a weighted, least-square best-fit PC relation:

\begin{equation}
[3.6]-[8.0] = 0.039 \left[ \log(P)-1.0 \right] [\pm0.008] -
0.019[\pm0.011] 
\end{equation}

The only apparent outliers are some nearby Cepheids observed in full
frame mode that have large photometric errors, and in particular
Polaris which is consistently red in all IRAC colors by $\sim 0.1$
mag. This detection is above the PSF-fitting photometry uncertainty
for this source, but should be taken with care, given that this
observation was not planned to maximize the accuracy of the PSF fit,
and may suffer from uncharacterized systematic errors. The large
scatter in the $[3.6]-[4.5]$ and $[3.6]-[5.8]$ color (as large as
$\sim 0.1$~mag) prevents the determination of a meaningful PC relation
in these colors. This scatter (which tends to be larger for long
period Cepheids) may be related to either some intrinsic property of
the stellar emission (the presence of broad spectral features in their
mid-IR spectrum) or some external cause (emission/absorption features
of interstellar or circumstellar origin). These two possibilities will
be examined in detail in the next section.

Colors combining IRAC and MIPS bands have an additional scatter due to
the different epochs at which the data were acquired. We can estimate
this scatter to be at least $\pm 0.4$~mag, as determined in the
previous section for the IRAC bands (assuming that the variability at
MIPS wavelengths is negligible, which is not necessarily the case). To
be detected, any excess in the $[3.6]-[24]$ and $[3.6]-[70]$ colors
(or other combinations mixing IRAC and MIPS bands) must be larger than
0.4~mag. Figure~\ref{fig-pc} shows that none of our sources have
excess in the $[3.6]-[24]$ color above this threshold (dashed
line). None of the sources with a 70~\micron{} measured flux has a
$[3.6]-[70]$ excess above $3 \sigma$ of the minimum expected
variability scatter (dotted line), with only sources undetected at
70~\micron{} above that line. Similar results are obtained by
combining the MIPS 24 and 70~\micron{} bands with the other IRAC
channels.

The $[24]-[70]$ color does not suffer from multi-epoch scatter, but
the lack of good photometric measurements for the 70~\micron{} band
prevent us from determining reliable colors for most of the
sources. Only one target (SZ~Tau) has a non-zero $[24]-[70]$ color at
more than 3$\sigma$ significance, which may indicate the presence of a
true color excess from this star. The significance of this detection
will be described in the next paper of this series (Barmby et al., in
preparation).

\section{Cepheid Mid-IR Colors and Search for Infrared
  Excess}\label{sec-excess}

Dusty mass loss in the wind of a Cepheid star would reveal itself in
the form of an infrared excess caused by thermal radiation from the
dust grains. IRAC and MIPS are particularly sensitive to this excess,
as shown in the case of Asymptotic Giant Branch (AGB) stars
\citep{marengo2008}. To isolate any excess from our data, we need
however to first characterize the intrinsic colors of Cepheids at
mid-IR bands. While a systematic study of the infrared spectral
properties of Classical Cepheids is beyond the scope of this work, we
can use available models of the 10~d period Cepheid $\zeta$~Gem (one
of our target stars) computed by \citet{marengo2002} as a test case.

\citet{marengo2002} developed a method to derive time-dependent
hydrodynamic models of classical Cepheids by solving Local
Thermodynamical Equilibrium (LTE) plane parallel radiative transfer
for a dynamic stellar atmosphere. The dynamic atmosphere, computed in
spherical geometry and in non-LTE conditions, was in turn driven by a
``pulsation piston'' reproducing the correct radial profile of the
star. While this procedure was not simultaneously solving the
hydrodynamic and radiative transfer solution for the star, it was
still able to produce realistic time-dependent intensity spectra, then
used to evaluate the temporal variations of the stellar limb darkening
at optical and infrared wavelengths \citep{marengo2003}. The method
was applied to the case of $\zeta$~Gem to estimate the effect of limb
darkening variations on the star's distance determination with the
geometric Baade-Wesselink method \citep{marengo2004}.

\begin{figure}[h!]
\centering
\includegraphics[width=0.6\textwidth, angle=0]{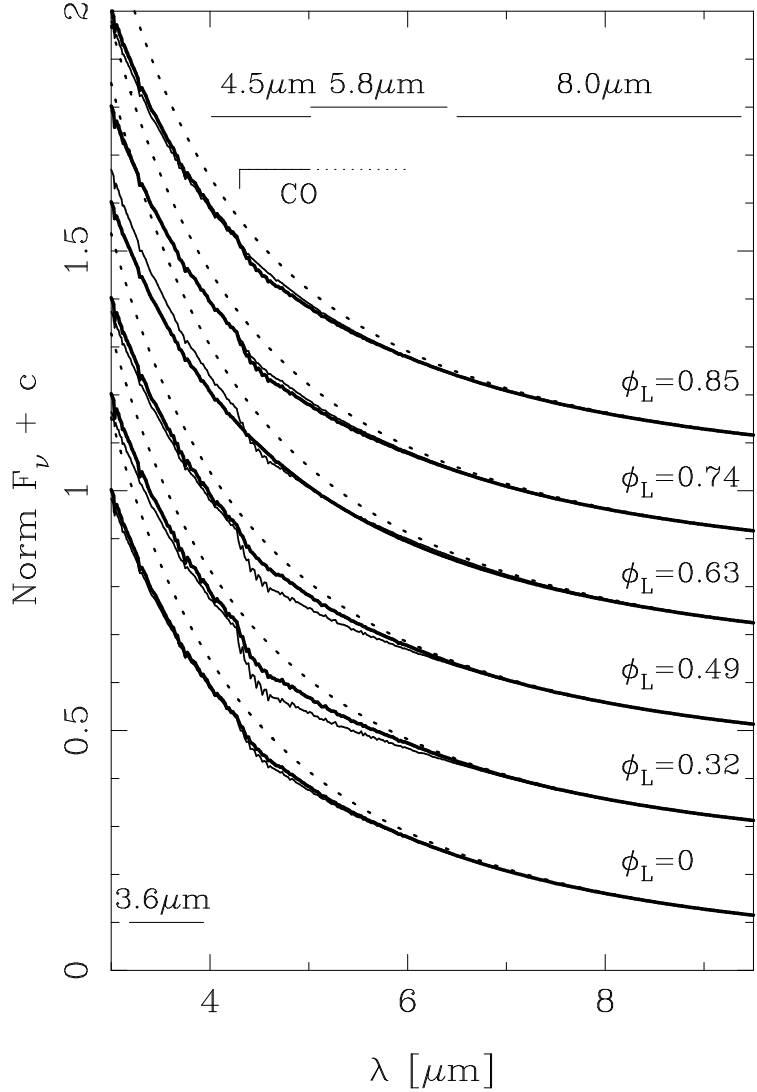}
\caption{Mid-IR model spectra of the 10~d period Cepheid
  $\zeta$~Gem. The thick solid lines are time-dependent hydrodynamic
  models derived by \citet{marengo2002}. The thin solid lines are
  instead equivalent models having the same $T_{eff}$ and
  $\log g$ as the dynamic models, but an atmosphere in hydrostatic
  equilibrium. The dotted lines are Rayleigh-Jeans blackbody fits
  of the spectra for $\lambda \ga 8$~\micron{}. The spectra are
  plotted for a number of significant phases: maximum V luminosity
  ($\phi_L=0$), maximum radius ($\phi_L=0.32$), minimum V luminosity
  ($\phi_L=0.49$), phase at which a shock-wave is crossing the
  photosphere ($\phi_L=0.63$), minimum radius ($\phi_L=0.74$) and the
  phase at which the star was observed with IRAC
  ($\phi_L=0.85$). A broad CO spectral feature is present at most
  phases, but with different strength, within the 4.5 and
  5.8~\micron{} IRAC pass-bands.}\label{fig-spectra}
\end{figure}

Figure~\ref{fig-spectra} shows the hydrodynamic model spectra of
$\zeta$~Gem at 5 significant phases: $\phi_L=0$ (maximum visible
luminosity), $\phi_L=0.32$ (maximum radius), $\phi_L=0.49$ (minimum
visible luminosity), $\phi_L=0.63$ (a phase at which a shock-wave is
crossing the photosphere), $\phi_L=0.74$ (minimum radius) and
$\phi_L=0.85$ (the phase at which the star was observed with IRAC). The
thick line shows the hydrodynamic model spectra, while the thin line
shows an equivalent hydrostatic equilibrium atmosphere having the same
$T_{eff}$ and $\log g$ of the dynamic models. The two sets of spectra
are compared with Rayleigh-Jeans black bodies normalized to the model
spectral emission for $\lambda \ga 8$~\micron.

It is apparent from the plot that the model spectra of $\zeta$~Gem
strongly depart from black body spectra at IRAC wavelengths: we should
thus expect IRAC colors that are significantly different from zero
magnitudes. The second most obvious characteristic of the models is
the presence of a broad spectral feature between 4 and 6~\micron{},
due to CO molecular absorption. This feature falls within the
band-passes of the IRAC 4.5 and 5.8~\micron{} filters, and is strong
enough to have a significant effect on the IRAC colors including these
bands. This feature is present in both the dynamic and static models,
and is thus a general property of stars with the temperature and
gravity of Cepheids, rather than being a consequence of the
pulsations. The strength of the feature appears however to be variable
with phase: it is stronger when the stellar atmosphere is more
expanded, as at maximum radius. When the atmosphere is compressed (as
at minimum radius, or when a shock-wave is crossing the photosphere in
the dynamic models), the feature is reduced to a minimum. The
absorption is also generally stronger in the static than in the
dynamic model spectra (the former corresponding to an atmosphere with
a more ``relaxed'' structure, as opposed to the ``compressed''
atmosphere in the dynamic models). This is not the only difference
between the static and dynamic models: at $\phi_L = 0.63$ (when a
shock-wave is transiting in the continuum-forming photosphere) the
dynamic and static atmospheres have very different continuum slopes in
their spectral emission, reflecting the dramatic consequences of
the pulsations in the affecting the structure of the Cepheid
atmosphere.

\begin{figure}[h!]
\centering
\includegraphics[width=0.65\textwidth, angle=-90]{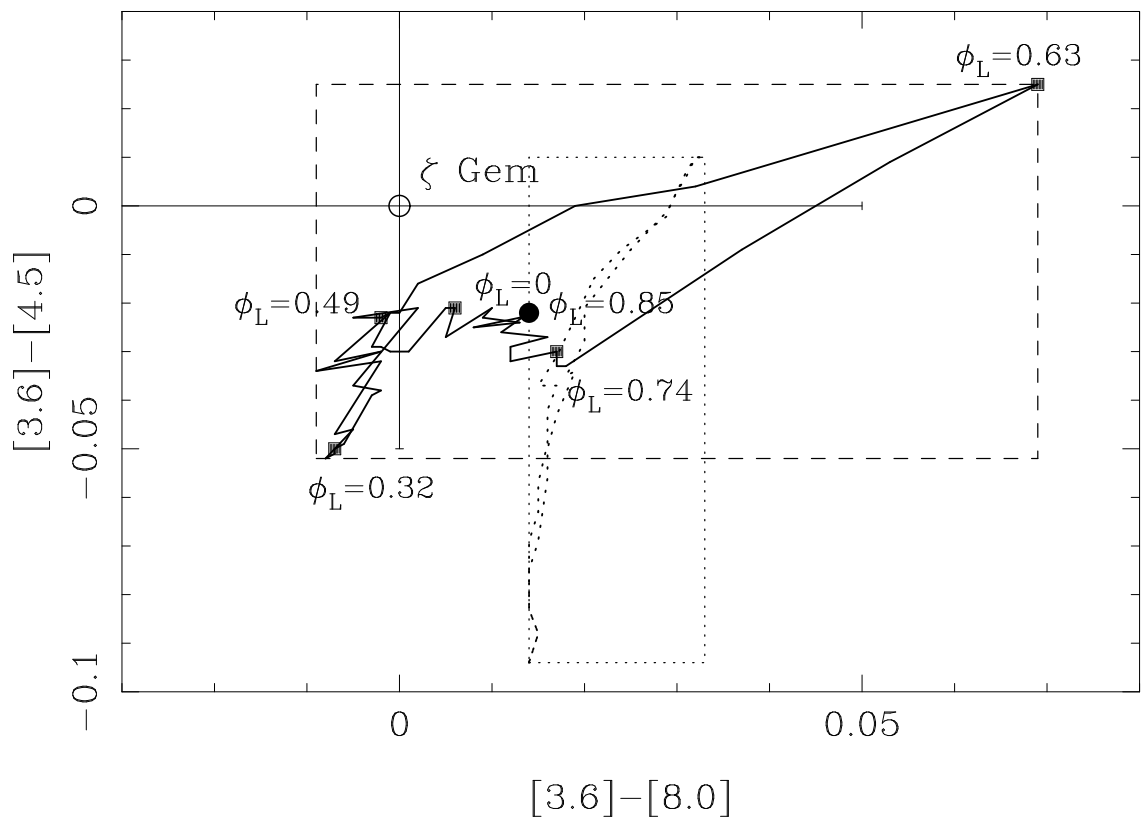}
\caption{IRAC colors of the dynamic (solid line) and static (dotted
  line) models of $\zeta$~Gem. The jitter on the model tracks is due
  to the finite resolution of the $T_{eff}$ and $\log g$ model
  grid. The colors of the dynamic model at the same phases of the
  spectra plotted in Figure~\ref{fig-spectra} are marked (square
  points). The model and actual colors of the source at the epoch of
  the IRAC observations are also plotted. The dashed and dotted square
  boxes indicate the ``bounding boxes'' of the two class of models
  (dynamic and static respectively) in the IRAC color
  space.}\label{fig-modcol}
\end{figure}

The presence of this CO absorption, and of the time-dependent
hydrodynamic effects, induce significant color variations in the IRAC
bands. Figure~\ref{fig-modcol} shows the synthetic IRAC colors of the
$\zeta$~Gem dynamic and static models. The largest spread is in the
$[3.6]-[4.5]$ color ($\sim 0.8$~mag and $\sim 1$~mag in the dynamic
and static model respectively), while the $[3.6]-[8.0]$ color shows a
much smaller variation ($\sim 0.08$~mag and $\sim 0.02$~mag). The
reason for this difference is that the $[3.6]-[4.5]$ color is affected
by the variable CO absorption, while the $[3.6]$ and $[8.0]$ bands are
outside this spectral feature. The dynamic model colors in fact are
confined for most phases in a smaller region ($\sim 0.03$~mag in both
colors), with the exception of a large excursion when the shock-wave
crossing the $\zeta$~Gem photosphere at $\phi_L = 0.63$ suddenly
expands the atmosphere. The $[3.6]-[4.5]$ colors of the static models
have instead a large variation that follows closely the changes in the
effective temperature of the star.

The figure also shows the actual colors of $\zeta$~Gem as observed
with IRAC. Due to the large uncertainty of this star's photometry, the
measured colors cannot be used to assess the respective accuracy of
the two class of models (the measurement is within 1$\sigma$ of both
model colors at the observation's phase $\phi_L = 0.85$).

\begin{figure}[h!]
\centering
\includegraphics[width=0.65\textwidth, angle=-90]{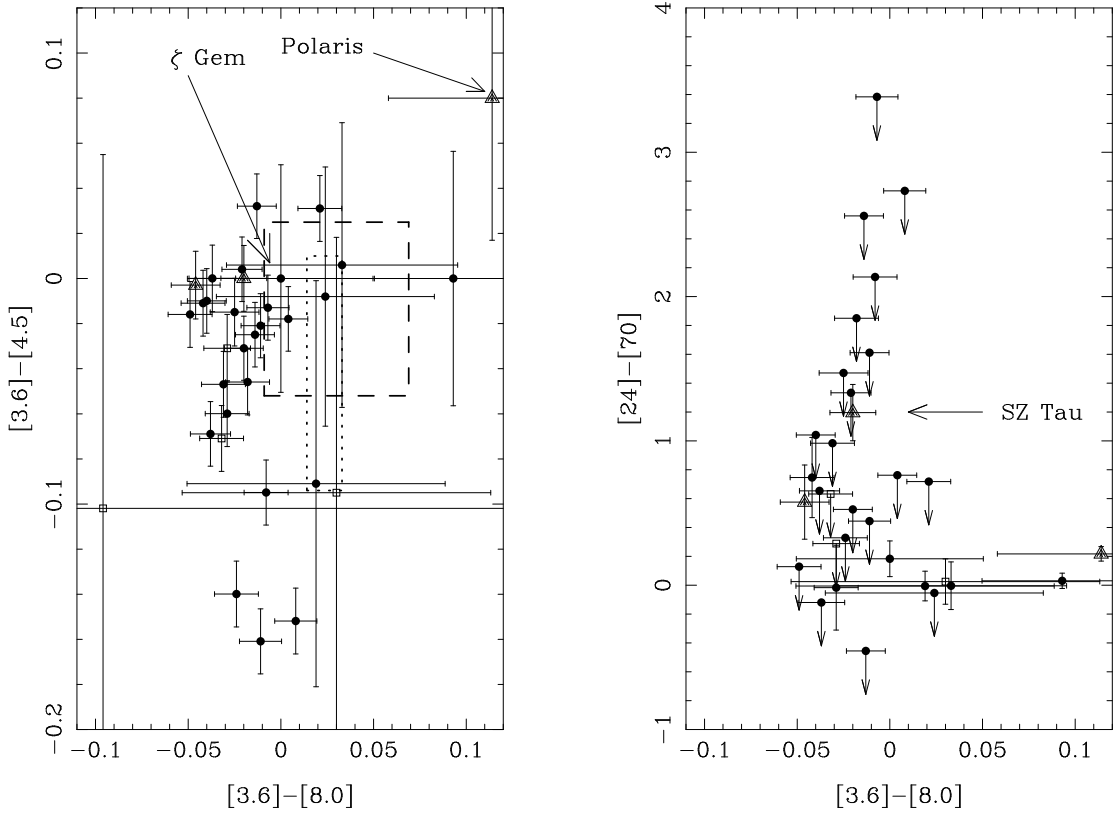}
\caption{Color-color diagrams of our sample. \emph{Left panel} ---
  IRAC $[3.6]-[8.0]$ vs. $[3.4]-[4.5]$ colors. The $[3.6]-[8.0]$ color
  shows a small spread ($\sim 0.05$~mag) for most of the sources. The
  $[3.6]-[4.5]$ color has instead a much larger spread, well above the
  variations allowed by the photometric uncertainties. Polaris is the
  source with redder colors ($\sim 0.1$~mag). The dashed and dotted
  boxes indicate the location of the $\zeta$~Gem dynamic and static
  models, respectively. \emph{Right panel} --- IRAC $[3.6]-[8.0]$
  vs. MIPS $[24]-[70]$ color: only one source (SZ~Tau, triangle
  indicated by arrow) has a statistically significant excess in the
  $[24]-[70]$ color. Other sources seem to be on a sequence of
  increasing colors, but the trend may be a spurious artifact, due to
  the absence of reliable 70~\micron{} detections. The 3 supergiants
  and the red giant in our sample (square symbols) have
  indistinguishable colors from those of the Cepheid stars, in both
  diagrams.}\label{fig-col}
\end{figure}

Figure~\ref{fig-col} shows the IRAC and MIPS color-color diagrams of
the program stars (including the 3 non-variable supergiants and 1 red
giant added to the sample for comparison). The dashed and dotted boxes
indicate the location of the dynamic and static model IRAC colors. It
is clear that the color variation of the $\zeta$~Gem models are
sufficient to explain the color spread present for most of the program
stars. The dynamic models are actually closer in both axes to the
actual colors of the observed stars, while the static models seem to
have a systematic offset of $\sim 0.04$~mag in the $[3.6]-[8.0]$ color
(the stars appear to be bluer). While it is unwise to extract general
considerations from this effect (the color difference is barely
significant given the photometric uncertainty, and the models are
specific to only one of the stars), this discrepancy suggests that the
color effects induced by the dynamics of the Cepheid pulsations may be
important in determining their actual IRAC colors. Three stars have a
$[3.6]-[4.5]$ colors ($\la -0.15$) significantly different from the
range predicted by both the dynamic and static models. These stars are
among the longest period Cepheids in our sample. This suggest that
long period Cepheids may present a deeper CO absorption, at least at
some pulsation phases, than $\zeta$~Gem. Shorter period Cepheids
instead show small $[3.6]-[4.5]$ color scatter, suggesting less
variability in the CO feature. Polaris shows $\sim 0.1$~mag
redder colors than all other sources or models, but this needs to be
confirmed with higher precision photometry for this star.

The colors of the 4 non-variable control stars are indistinguishable
from the colors of the Cepheids, confirming that the mid-IR CO
spectral feature that is responsible for the color shift is common in
giant stars with this spectral type even when not on the Cepheid
instability strip.

The comparison of the observed IRAC colors with the $\zeta$~Gem
models spectra indicates that the $[3.6]-[4.5]$ and $[3.6]-[5.8]$
scatter in the PC plots in Figure~\ref{fig-pc} is intrinsic. The
$[3.6]-[8.0]$ colors are consistent with the model colors. We
therefore conclude that there is no measurable extrinsic infrared
excess in the IRAC bands, with only the possible exception of
Polaris. Given that $[3.6]-[8.0]$ excess would correspond to the
presence of warm dust ($T_d \sim 500$~K), this means that we do not
find compelling evidence, for most sources, of circumstellar
\emph{dusty} wind that could be associated with currently active
\emph{dusty} mass loss (contrary to what was stated in
\citealt{marengo2009b} where Cepheid intrinsic color variations were
not taken into account). This result does not exclude however
the presence of winds with low dust content.

The $[24]-[70]$ color does not suffer from variability-induced
scatter, but the lack of good photometric measurements for the
70~\micron{} band prevents determining reliable colors for most
of the sources. The plot shows a trend in this color for the stars in
the sample, but this trend is most likely a consequence of having only
upper limits for the 70~\micron{} brightness of most stars. Only one
star (again SZ~Tau) has a $[24]-[70]$ excess above its 3$\sigma$
photometric error, which may indicate the presence of cold dust ($T_d
\sim 50$~K) in the proximity of the star.

\section{Discussion}\label{sec-discussion}

The reliability of Classical Cepheids as standard candles is of
paramount importance for astronomy. The recent advances in infrared
space astronomy have shifted the focus of obtaining accurate PL
relations to wavelengths longer than the visible, where interstellar
extinction is reduced. \citet{madore2009b} and \citet{freedman2009}
have demonstrated how PL relations obtained at IRAC wavelengths can be
effectively used to measure the distance of nearby galaxies. This work
is a further step in the characterization of PL relations in all
\spitzer{} photometric bands, including two MIPS bands at 24 and
70~\micron, by using a sample of Classical Cepheids in the Galaxy.

In order to provide a detailed comparison between theory and
observations we adopted the large set of nonlinear, convective models
computed by \citet{bono1999, marconi2005} and by
\citet{fiorentino2007}. For Galactic Cepheids we chose a scaled-solar
chemical composition (helium, $Y=0.28$; metals, $Z=0.02$) and
accounted for fundamental mode pulsators. Moreover, we covered a broad
range of stellar masses ($3.5 \le M/M_\odot \le 11.5$), and to account
for current uncertainties affecting the size of the helium core
\citep{bono2006} we adopted two different mass-luminosity relations
based on canonical and non-canonical (the latter incorporating a range
of main sequence core convective overshoot) evolutionary
models. Theoretical predictions were transformed into the
observational plane using scaled-solar atmospheres based on {\tt
  ATLAS9} models \citep{castelli2003}, and multiplied with the IRAC
band-passes. The slopes and zero points of these model Leavitt laws
are listed in the last two columns of Table~\ref{tab-PL}. While the
zero points are in general agreement with all our fits, the slopes are
significantly shallower than our best fits obtained with both ``new''
and ``old'' distances (the disagreement is however reduced in the fits
adopting the ``new'' distances). The PL relations derived with the
exclusive use of the astrometric distances (that do not rely on the
$p(P)$ relation), are however in excellent agreement with the
theoretical relations. This result may suggest that the period
dependence of the $p-$factor currently adopted in IRSB distances may
still need further refinement. Because of the small number of Cepheids
and the limited range of periods for which accurate parallaxes are
available, we could not attempt to invert the problem and estimate the
$p(P)$ dependence with our data.

Comparison between our Galactic PL relations and the relations derived
for the LMC by \citet{madore2009a} and \citet{ngeow2009} leads to
contradictory results. The PL relations obtained with both the ``new''
and ``old'' IRSB distances are steeper than the LMC relations, while
the Galactic PL relations we obtain with the astrometric distances are
significantly shallower. To add to the confusion, it should be noted
that the slopes derived by \citet{ngeow2009} and \citet{madore2009a}
disagree by more than their respective uncertainties, despite using
stars from the same galaxy\footnote{The difference between the two LMC
  results can be an issue of crowding and binarity in the
  \citet{ngeow2009} sample (selected without individual image
  inspection, and containing a larger fraction of fainter short
  period Cepheids) that could result in shallower slopes.}. The
contradictory results in our Galactic PL relation slopes shows once
more the effects of the systematics introduced by the IRSB distances
and their dependence on the $p(P)$ relation. Based on the more
reliable astrometric distances we should conclude that the slope of
the PL relation could be shallower at Galactic metallicity than in the
LMC, but even this result is not statistically significant, due to the
larger uncertainty in the PL slope. In conclusion, the insufficient
reliability of the IRSB distances, and the small number of stars for
which astrometric distances are available, prevents us from resolving
the dependence of the PL relations from metallicity. This problem has
already been noticed at optical and near-IR wavelengths by other
authors (see e.g. \citealt{storm2004, fouque2007,
  romaniello2008}). \citet{riess2009} also noted that the precision of
current datasets does not allow measurement of the dependence of the
PL relation on metallicity; those authors measured statistically
consistent slopes at 1.6 micron for the Milky Way, the LMC and other
galaxies. 

We also cannot determine a reliable wavelength dependence of the PL
slope, as all our values are within 1$\sigma$ from the average slope
value. This is consistent with the expected flattening of the PL slope
wavelength dependence in the mid-IR shown by the models. If we compare
our slope values with Figure~4 in \citet{freedman2008}, however, we
note a similar trend. Our result for 4.5~\micron{} is marginally
higher (by $\sim 0.1$) than the slope at the 3.6 and 8.0~\micron, as
also shown by the LMC Cepheid fit (and by the models). This anomaly in
the 4.5~\micron{} (and, to a minor extent, 5.8~\micron) slope is
related to the presence of the variable CO band, which increases the
amplitude variation in these bandpasses.

For the first time we have measured the PL relation in the MIPS 24 and
70~\micron{} bands. The 24~\micron{} relation, in particular, is
important for the determination of Cepheid distances in the Galactic
plane from datasets such as MIPSGAL \citep{carey2009}, where high
extinction from ISM dust may complicate their detection at IRAC bands. 

The determination of a PC relation is in principle not limited by the
uncertainty in the distances affecting our PL
relations. Figure~\ref{fig-pc}, however, clearly shows that such a
relation can be derived with sufficient accuracy only in the
$[3.6]-[8.0]$ color. Other colors show a scatter that is increasingly
larger with the period, as high as $\sim 0.1$~mag, one order of
magnitude larger than our color accuracy of $\sim 0.02$. The fact that
this scatter was not found by \citet{ngeow2009} may again be a
consequence of the difference in the median period between our
Galactic and their LMC sample.

Based on our data and on numerical modeling of one representative
Cepheid ($\zeta$~Gem), we conclude that the color scatter affecting
long period Cepheids is a consequence of variable stellar CO
absorption, rather than infrared excess from circumstellar
dust. Circumstellar dust at $\sim 500$~K (the temperature where dust
thermal emission is maximum in the IRAC bands) would produce the
largest excess in the $[3.6]-[8.0]$ color, which we do not observe,
with the possible exception of a weak detection ($\sim 0.1$~mag)
around Polaris. Our $\zeta$~Gem models show that scatter in the IRAC
4.5 and 5.8~\micron{} bands can be induced by variations in effective
temperature and/or the propagation of shocks through the Cepheid
atmospheres. A more comprehensive modeling effort, including Cepheids
with different periods and pulsation modes, is however required to
confirm and quantify this effect.

Not finding an infrared excess in most, if not all, stars is somewhat
unexpected, given the detection of circumstellar shells by near-IR
interferometers. Figure~8 in \citet{merand2007} shows a number of
targets, including the long period Cepheids Y~Oph and $\ell$~Car (both
part of our sample), with a significant (4--5\%) excess in the K
band. We do not measure any excess of this magnitude in these or other
stars, which seems to rule out the presence of circumstellar dust in
the shells detected around these and other Cepheids. The
interferometric determination of the radius of these circumstellar
shells, however, indicate values as low as $\sim 2$ stellar radii
\citep{merand2006}. For a Cepheid with $T_{\textrm{eff}} \sim
5$,000--6,000~K, dust equilibrium temperature at $\sim 2$~R$_*$ would
be $T_d \simeq T_*/\sqrt{R_d/R_*} \la 3$,500~K. This is well above the
sublimation temperature of any known astronomical dust.

Our results can be reconciled with the interferometric detection of
circumstellar shells close to the star if the emission is not due to
dust, but rather to some strong molecular line emission. Given that
the observations in \citet{merand2006, merand2007} and
\citet{kervella2006} were made in the $K$ band, a possible
candidate for this emission is shocked $H_2$. The presence of $H_2$
shocked emission lines could further contribute to the scatter in the
IRAC 4.5~\micron{} band, for certain densities and shock velocities
\citep{smith2006}. This hypothesis supports the idea that Cepheid
stars indeed have a strong stellar wind associated with a pulsation
driven mass-loss mechanism, as suggested by \citet{merand2007}. Unlike
the case of mass loss in red giants, AGB and supergiant stars (showing
a dust-driven wind), this stellar wind would be largely dustless, due
to the higher temperature of the star. This hypothesis needs to be
tested by near-IR multi-epoch spectral monitoring of these stars.

This hypothesis does not imply that a Cepheid stellar wind is
completely devoid of dust, that may be condensing at larger radii, in
quantities below the detection limits of our IRAC observations (see
$\ell$~Car mid-IR detection of mid-IR circumstellar emission with the
VLT MIDI and VISIR instruments, \citealt{kervella2009}). Cold dust may
also be collected by the outflow, at much larger distances (thousands
of AU), from the ISM, as in the case of the well known nebula around
RS~Pup \citep{kervella2009}. We find evidence supporting this
mechanism in our IRAC and MIPS images. Extended emission at 5.8, 8.0,
24 and 70~\micron{} around $\delta$~Cep indicate that this star may be
losing mass due to a strong wind pushing into the local interstellar
medium, which is leading to the formation of a 70~\micron{} bow shock
detected at large distance from the star ($\sim 10$,000~AU). A
detailed analysis of this phenomenon is being published elsewhere
(Marengo et al., in preparation).  In a more comprehensive paper
\citep{barmby2009} we will discuss the presence of spatially resolved
extended emission at 24 and 70~\micron{} around more targets, and
quantify the occurrence of mass loss in the Cepheid phase that can be
inferred from our imaging data.

\section{Summary}\label{sec-summary}

We have derived the PL and PC relations in \spitzer/IRAC bands for a
sample of Galactic Cepheids. These relations are critically dependent
on the choice of the period dependence of the $p$-factors used for the
distance determination with the IRSB method, even though the
uncertainties in our fits prevent the assessment of the best $p(P)$
relation choice. The best agreement with theoretical PL relations,
however, is obtained when only distances obtained by astrometric
methods are used. We do not detect statistically significant
variations between the slope and zero points of the PL relations
between our Galactic sample and LMC relations obtained by
\citet{madore2009a}, despite the difference in metallicity. We find
that the intrinsic variations in the 4.5 and 5.8~\micron{} fluxes are
larger for long period Cepheids. These variations (of the order of
$\sim 0.1$~mag) are related to deep CO absorption, dependent on the
stellar $T_{eff}$ and hydrodynamic effects associated to the stellar
pulsations. We do not find significant infrared excess related to warm
circumstellar dust, except for a weak excess detected at IRAC
wavelengths for Polaris, and at 70~\micron{} for SZ~Tau. This may rule
out the presence of extensive dust driven mass loss in the Cepheid
phase, but leaves open the possibility of pulsation-driven mass loss
from a dust-poor wind, as suggested by recent interferometric
observations.



\acknowledgments

This work is based on observations made with the \spitzer{} Space
Telescope, which is operated by the Jet Propulsion Laboratory,
California Institute of Technology under NASA contract 1407. P.~B. and
D.~W. both acknowledge research support through Discovery Grants from
the Natural Sciences and Engineering Research Council of
Canada. N.~R.~E. acknowledges support from the Chandra X-Ray Center
grant NAS8-03060. We thank Robert Kurucz and Chow-Choong Ngeow
for helpful discussions during the preparation of this manuscript.  We
also acknowledge M. Marconi and P.G. Prada Moroni for sending us the
predicted Cepheid PL relations in advance of publication.



{\it Facilities:} \facility{Spitzer (IRAC, MIPS)}.





\clearpage



\begin{deluxetable}{lccccccc}
\tabletypesize{\footnotesize}
\rotate
\tablewidth{0pt}
\tablecaption{Target list and catalog of observations}
\tablehead{
  \colhead{Name} &
  \colhead{IRAC sub JD\tablenotemark{a}} &
  \colhead{IRAC full JD\tablenotemark{a}} &
  \colhead{MIPS JD\tablenotemark{a}} &
  \colhead{$D$ [kpc]\tablenotemark{b}} &
  \colhead{$D$ [kpc]\tablenotemark{c}} &
  \colhead{$P$ [d]} &
  \colhead{Mode}}
\startdata
\objectname[DT Cyg]{DT~Cyg}   
             & 54065.535 & \nodata   & 53935.208 
             & 0.457$\pm$0.069   &  0.535$\pm$0.029\tablenotemark{d}
             & 2.54   & 1st \\
\objectname[SZ Tau]{SZ~Tau}   
             & 54006.076 & 54189.135 & 54400.742 
             & 0.513$\pm$0.029   &  0.556$\pm$0.034\tablenotemark{e}
             & 3.15   & 1st \\
\objectname[RT Aur]{RT~Aur}
             & 54036.201 & \nodata   & 54202.804 
             & 0.417$\pm$0.033\tablenotemark{f}
             & 0.417$\pm$0.033\tablenotemark{f}  & 3.73   
             & F   \\
\objectname[alf UMi]{Polaris}\tablenotemark{i}
             & \nodata   & 52991.065 & 54024.680 
             & 0.130$\pm$0.002\tablenotemark{f}
             & 0.130$\pm$0.002\tablenotemark{f}   &  3.97   
             & 1st \\
\objectname[BF Oph]{BF~Oph}
             & 53997.204 & \nodata   & 53979.664 
             & 0.637$\pm$0.032   &  0.714$\pm$0.011 & 4.07   
             & F   \\
\objectname[FF Aql]{FF~Aql}\tablenotemark{i}
             & 54039.420 & \nodata   & 54023.113 
             & 0.356$\pm$0.023\tablenotemark{f}
             & 0.356$\pm$0.023\tablenotemark{f}   &  4.47   
             & F   \\
\objectname[V350 Sgr]{V350~Sgr}\tablenotemark{i}   
             & 54002.135 & \nodata   & 54013.655 
             & 0.935$\pm$0.052   &  0.979$\pm$0.054\tablenotemark{d}
             & 5.15   & F   \\
\objectname[del Cep]{$\delta$~Cep}
             & \nodata   & 53957.101 & 53935.018 
             & 0.273$\pm$0.011\tablenotemark{f}   
             & 0.273$\pm$0.011\tablenotemark{f}   &  5.37   
             & F   \\
\objectname[V Cen]{V~Cen}
             & 53957.367 & 53957.365 & 53980.875 
             & 0.606$\pm$0.044   &  0.684$\pm$0.020  &  5.49
             & F   \\
\objectname[BB Sgr]{BB~Sgr}
             & 54002.629 & \nodata   & 54013.643 
             & 0.787$\pm$0.031   &  0.801$\pm$0.010  &  6.64
             & F   \\
\objectname[U Sgr]{U~Sgr}
             & 53999.677 & 53997.790 & 54009.048 
             & 0.565$\pm$0.019   &  0.585$\pm$0.006  &  6.75
             & F   \\
\objectname[V636 Sco]{V636~Sco}\tablenotemark{i}   
             & 53997.194 & \nodata   & 53983.521 
             & \nodata   & 0.839$\pm$0.046\tablenotemark{d}
             & 6.80   & F   \\
\objectname[U Aql]{U~Aql}\tablenotemark{i}   
             & 54039.442 & \nodata   & 54016.545 
             & 0.606$\pm$0.029   &  0.691$\pm$0.038\tablenotemark{d}
             & 7.02   & F   \\
\objectname[eta Aql]{$\eta$~Aql}
             & \nodata   & 54039.439 & 54238.612 
             & 0.241$\pm$0.014   &  0.250$\pm$0.005 & 7.18
             & F   \\
\objectname[W Sgr]{W~Sgr}\tablenotemark{i}
             & 53997.264 & \nodata   & 54013.542 
             & 0.439$\pm$0.038\tablenotemark{f}   
             & 0.439$\pm$0.038\tablenotemark{f}   &  7.59   
             & F   \\
\objectname[GH Lup]{GH~Lup}
             & 53961.333 & \nodata   & 53979.733 
             & 1.124$\pm$0.203   &  1.023$\pm$0.051\tablenotemark{g} & 9.28   
             & F   \\
\objectname[S Mus]{S~Mus}\tablenotemark{i}   
             & 53960.571 & \nodata   & 53935.332 
             & 0.820$\pm$0.054   &  0.916$\pm$0.046\tablenotemark{g} & 9.66   
             & F   \\
\objectname[S Nor]{S~Nor}
             & 53999.638 & 54351.249 & 53983.802 
             & 0.943$\pm$0.044   &  0.959$\pm$0.014  & 9.75
             & F   \\
\objectname[bet Dor]{$\beta$~Dor}
             & \nodata   & 53960.988 & 53935.322 
             & 0.318$\pm$0.016\tablenotemark{f}   
             & 0.318$\pm$0.016\tablenotemark{f}   &  9.84   
             & F   \\
\objectname[zet Gem]{$\zeta$~Gem}
             & \nodata   & 54228.830 & 54046.15 
             & 0.360$\pm$0.023\tablenotemark{f}   
             & 0.360$\pm$0.023\tablenotemark{f}   & 10.15   
             & F   \\
\objectname[X Cyg]{X~Cyg}
             & 54065.539 & \nodata   & 53935.023 
             & 1.163$\pm$0.027   & 1.213$\pm$0.010 & 16.39
             & F   \\
\objectname[Y Oph]{Y~Oph}
             & 53997.328 & \nodata   & 54202.088 
             & 0.552$\pm$0.040   
             & 0.573$\pm$0.008\tablenotemark{h} & 17.13   
             & F   \\
\objectname[VY Car]{VY~Car}
             & 53957.067 & \nodata   & 53935.360 
             & 1.818$\pm$0.099   & 1.994$\pm$0.020 & 18.91
             & F   \\
\objectname[SW Vel]{SW~Vel}
             & 54098.538 & \nodata   & 54202.474 
             & 2.381$\pm$0.057   & 2.506$\pm$0.029 & 23.44
             & F   \\
\objectname[T Mon]{T~Mon}\tablenotemark{i}
             & 54395.313 & \nodata   & 54021.550 
             & 1.389$\pm$0.058   & 1.455$\pm$0.037 & 27.02
             & F   \\
\objectname[AQ Pup]{AQ~Pup}
             & 54228.851 & \nodata   & 54107.409 
             & 3.030$\pm$0.184   & 3.194$\pm$0.066 & 30.10
             & F   \\
\objectname[l Car]{$\ell$~Car}
             & \nodata   & 53960.984 & 53935.328 
             & 0.498$\pm$0.050\tablenotemark{f}   
             & 0.498$\pm$0.050\tablenotemark{f}   & 35.55   
             & F   \\
\objectname[U Car]{U~Car}
             & 53960.567 & \nodata   & 53935.349 
             & 1.492$\pm$0.067   & 1.565$\pm$0.023 & 38.77
             & F   \\
\objectname[RS Pup]{RS~Pup}
             & 54098.477 & \nodata   & 54073.425 
             & 1.818$\pm$0.099   & 2.052$\pm$0.151 & 41.39
             & F   \\
\objectname[HD183864]{HD183864}\tablenotemark{j}   
             & 54064.419 & \nodata   & 54046.098 
             & \nodata & \nodata & \nodata & \nodata \\
\objectname[HD182296]{HD182296}\tablenotemark{j}   
             & 54226.294 & \nodata   & 54018.537 
             & \nodata & \nodata & \nodata & \nodata \\
\objectname[gam Phe]{$\gamma$~Phe}\tablenotemark{j}   
             & 53957.014 & \nodata   & 53935.315 
             & \nodata & \nodata & \nodata & \nodata \\
\objectname[psi And]{$\psi$~And}\tablenotemark{j}   
             & \nodata   & 54094.302 & 54106.889
             & \nodata & \nodata & \nodata & \nodata \\
\enddata
\tablenotetext{a}{JD-2400000.5 at observation start}
\tablenotetext{b}{Distance from \citet{fouque2007} unless noted}
\tablenotetext{c}{Distance from \citet{storm2004} unless noted}
\tablenotetext{d}{This work (Table~\ref{tab-extrad})}
\tablenotetext{e}{Distance from \citet{groenewegen2004}}
\tablenotetext{f}{Distance from \citet{benedict2007} or
 \citet{vanleeuwen2007} for Polaris}
\tablenotetext{g}{Distance from \citet{romaniello2008} (with 5\%
  assumed uncertainty)} 
\tablenotetext{h}{Distance from \citet{barnes2005}, Least Square fit}
\tablenotetext{i}{Spectroscopic binary}
\tablenotetext{j}{Supergiant or M giant ($\gamma$~Phe) star}
\label{tab-targets}
\end{deluxetable}

\clearpage



\begin{deluxetable}{lcccccccc}
\tabletypesize{\footnotesize}
\rotate
\tablewidth{0pt}
\tablecaption{``Old'' distances missing from previous catalogs}
\tablehead{
  \colhead{Name} &
  \colhead{$R_0 [R_\odot]$} &
  \colhead{$\theta_0$ [mas]} &
  \colhead{$V$} &
  \colhead{$K$} &
  \colhead{$E(B-V)$} &
  \colhead{$V_0$} &
  \colhead{$K_0$} &
  \colhead{$d$ [kpc]}}
\startdata
DT~Cyg   & 30.7$\pm$0.9    & 0.570$\pm$0.018 & 5.744$\pm$0.010
         & 4.430$\pm$0.010 & 0.039$\pm$0.020  & 5.635$\pm$0.010
         & 4.416$\pm$0.010 & 0.535$\pm$0.029 \\
V350~Sgr & 40.7$\pm$4.5    & 0.424$\pm$0.047 & 7.483$\pm$0.010
         & 5.141$\pm$0.010 & 0.312$\pm$0.020 & 6.516$\pm$0.010
         & 5.031$\pm$0.010 & 0.979$\pm$0.054 \\
V636~Sco & 50.0$\pm$4.8    & 0.568$\pm$0.054 & 6.654$\pm$0.010
         & 4.409$\pm$0.010 & 0.217$\pm$0.020 & 5.981$\pm$0.010
         & 4.332$\pm$0.010 & 0.839$\pm$0.046 \\
U~Aql    & 51.3$\pm$6.0    & 0.830$\pm$0.097 & 6.446$\pm$0.010
         & 3.893$\pm$0.010 & 0.399$\pm$0.020 & 5.209$\pm$0.010
         & 3.752$\pm$0.010 & 0.691$\pm$0.038 \\
\enddata
\label{tab-extrad}
\end{deluxetable}



\begin{deluxetable}{lcccccc}
\tabletypesize{\footnotesize}
\rotate
\tablewidth{0pt}
\tablecaption{IRAC and MIPS photometric calibration}
\tablehead{
  \colhead{Item} &
  \colhead{[3.6]} &
  \colhead{[4.5]} &
  \colhead{[5.8]} &
  \colhead{[8.0]} &
  \colhead{[24]} &
  \colhead{[70]}}
\startdata
Isophotal $\lambda$ [$\mu$m] & 3.550 & 4.493 & 5.731 & 7.872 & 23.68 & 71.42 \\
FLUXCONV [(MJy/sr)/(DN/s)]   & 0.1088 & 0.1388 & 0.5952 & 0.2021 & 0.0447 & 702. \\
6.1'' aperture corr. [mag]\tablenotemark{a}   & 0.051$\pm$0.002 & 0.055$\pm$0.002 & 0.056$\pm$0.006 & 0.065$\pm$0.005 & \nodata & \nodata \\
$f_{corr}$(subarray) & 0.994 & 0.995 & 0.939 & 1.006 & \nodata & \nodata \\
$F_\nu$(Vega) [Jy]           & 280.9$\pm$4.1 & 179.7$\pm$2.6 & 115.0$\pm$1.7 & 64.1$\pm$0.9 & 7.14$\pm$0.08 & 0.775$\pm$0.009 \\
\enddata
\tablenotetext{a}{Aperture used for FF~Aql, VY~Car and AQ~Pup
  IRAC subarray data} 
\label{tab-params} 
\end{deluxetable}

\clearpage



\begin{deluxetable}{lcrrrr}
\tabletypesize{\footnotesize}
\tablewidth{0pt}
\tablecaption{IRAC aperture vs. PSF fitting photometry
  comparison\tablenotemark{a}}
\tablehead{
  \colhead{Name} &
  \colhead{IRAC mode} &
  \colhead{[3.6]} &
  \colhead{[4.5]} &
  \colhead{[5.8]} &
  \colhead{[8.0]}}
\startdata
V Cen  & subarray   & 4.336$\pm$0.010 & 4.356$\pm$0.010 & 4.377$\pm$0.003 & 4.347$\pm$0.003 \\
''     & full frame & 4.332$\pm$0.270 & 4.392$\pm$0.156 & 4.392$\pm$0.093 & 4.362$\pm$0.060 \\
U Sgr  & subarray   & 3.811$\pm$0.010 & 3.856$\pm$0.010 & 3.864$\pm$0.008 & 3.829$\pm$0.006 \\
''     & full frame & 3.844$\pm$0.236 & 3.772$\pm$0.105 & 3.789$\pm$0.054 & 3.789$\pm$0.053 \\
S Nor  & subarray   & 3.969$\pm$0.010 & 4.015$\pm$0.010 & 4.035$\pm$0.008 & 4.000$\pm$0.006 \\
''     & full frame & 3.984$\pm$0.150 & 4.072$\pm$0.136 & 4.027$\pm$0.067 & 4.005$\pm$0.044 \\
SZ Tau & subarray   & 4.165$\pm$0.010 & 4.164$\pm$0.010 & 4.206$\pm$0.009 & 4.185$\pm$0.007 \\
''     & full frame & 4.144$\pm$0.200 & 4.144$\pm$0.200 & 4.221$\pm$0.080 & 4.195$\pm$0.076 \\
\tableline
Avg diff. & \nodata & $-$0.007$\pm$0.021 & 0.009$\pm$0.065 & 0.021$\pm$0.049 & 0.003$\pm$0.022 \\
\enddata
\tablenotetext{a}{With the exception of V~Cen, full frame and subarray
  data of each source have been obtained at different epochs, and the
  difference in photometry can be related not only on photometry
  biases, but also to the source variability}
\label{tab-test} 
\end{deluxetable}

\clearpage



\begin{deluxetable}{llrrrrlr}
\tabletypesize{\footnotesize}
\rotate
\tablewidth{0pt}
\tablecaption{Final \spitzer{} photometry of all targets}
\tablehead{
  \colhead{Name} &
  \colhead{IRAC mode} &
  \colhead{[3.6]} &
  \colhead{[4.5]} &
  \colhead{[5.8]} &
  \colhead{[8.0]} &
  \colhead{[24]} &
  \colhead{[70]}}
\startdata
    DT~Cyg & subarray     &  4.340$\pm$0.010 &  4.342$\pm$0.011 &  4.381$\pm$0.010 &  4.386$\pm$0.008 &  4.367$\pm$0.002 &  3.791$\pm$0.257 \\
    SZ~Tau & subarray     &  4.165$\pm$0.010 &  4.164$\pm$0.010 &  4.206$\pm$0.009 &  4.185$\pm$0.007 &  4.209$\pm$0.002 &  3.013$\pm$0.195 \\
    RT~Aur & subarray     &  3.865$\pm$0.010 &  3.880$\pm$0.010 &  3.924$\pm$0.008 &  3.914$\pm$0.006 &  3.773$\pm$0.002 &  3.645 \\
   Polaris & full frame   &  0.573$\pm$0.037 &  0.493$\pm$0.051 &  0.411$\pm$0.070 &  0.459$\pm$0.042 &  0.723$\pm$0.001 &  0.506$\pm$0.050 \\
    BF~Oph & subarray     &  5.141$\pm$0.010 &  5.136$\pm$0.010 &  5.171$\pm$0.005 &  5.162$\pm$0.004 &  5.031$\pm$0.005 &  3.698 \\
   FF~Aql & subarray     &  3.391$\pm$0.010 &  3.390$\pm$0.010 &  3.436$\pm$0.008 &  3.428$\pm$0.007 &  3.338$\pm$0.001 &  3.458 \\
  V350~Sgr & subarray     &  4.948$\pm$0.010 &  4.957$\pm$0.010 &  4.988$\pm$0.004 &  4.988$\pm$0.003 &  5.032$\pm$0.009 &  3.991 \\
$\delta$~Cep & full frame &  2.174$\pm$0.044 &  2.182$\pm$0.037 &  2.166$\pm$0.032 &  2.150$\pm$0.039 &  2.120$\pm$0.001 &  2.174 \\
     V~Cen & subarray     &  4.336$\pm$0.010 &  4.356$\pm$0.010 &  4.377$\pm$0.003 &  4.347$\pm$0.003 &  4.424$\pm$0.003 &  2.813 \\
    BB~Sgr & subarray     &  4.350$\pm$0.010 &  4.380$\pm$0.010 &  4.386$\pm$0.003 &  4.370$\pm$0.003 &  4.339$\pm$0.004 &  3.814 \\
     U~Sgr & subarray     &  3.811$\pm$0.010 &  3.856$\pm$0.010 &  3.864$\pm$0.008 &  3.829$\pm$0.006 &  3.760$\pm$0.003 &  1.910 \\
  V636~Sco & subarray     &  4.375$\pm$0.010 &  4.389$\pm$0.011 &  4.434$\pm$0.010 &  4.400$\pm$0.008 &  4.384$\pm$0.001 &  2.913 \\
     U~Aql & subarray     &  3.810$\pm$0.010 &  3.869$\pm$0.010 &  3.877$\pm$0.008 &  3.839$\pm$0.006 &  3.665$\pm$0.002 &  3.682$\pm$0.294 \\
$\eta$~Aql & full frame   &  2.003$\pm$0.041 &  1.997$\pm$0.048 &  1.990$\pm$0.021 &  1.970$\pm$0.047 &  1.856$\pm$0.001 &  1.860$\pm$0.165 \\
     W~Sgr & subarray     &  2.787$\pm$0.010 &  2.855$\pm$0.010 &  2.877$\pm$0.005 &  2.825$\pm$0.004 &  2.808$\pm$0.001 &  2.154 \\
    GH~Lup & subarray     &  4.692$\pm$0.010 &  4.716$\pm$0.010 &  4.731$\pm$0.004 &  4.706$\pm$0.003 &  4.644$\pm$0.006 &  2.085 \\
     S~Mus & subarray     &  3.835$\pm$0.010 &  3.845$\pm$0.010 &  3.884$\pm$0.008 &  3.877$\pm$0.006 &  3.902$\pm$0.001 &  3.156$\pm$0.278 \\
     S~Nor & subarray     &  3.969$\pm$0.010 &  4.015$\pm$0.010 &  4.035$\pm$0.008 &  4.000$\pm$0.006 &  4.086$\pm$0.002 &  3.102 \\
$\beta$~Dor & full frame  &  1.899$\pm$0.059 &  1.990$\pm$0.068 &  1.937$\pm$0.049 &  1.880$\pm$0.037 &  1.858$\pm$0.001 &  1.864$\pm$0.103 \\
$\zeta$~Gem & full frame  &  2.045$\pm$0.039 &  2.045$\pm$0.032 &  2.060$\pm$0.033 &  2.045$\pm$0.032 &  1.982$\pm$0.001 &  1.799$\pm$0.124 \\
     X~Cyg & subarray     &  3.741$\pm$0.010 &  3.880$\pm$0.010 &  3.841$\pm$0.007 &  3.765$\pm$0.006 &  3.680$\pm$0.001 &  3.352 \\
     Y~Oph & subarray     &  2.543$\pm$0.010 &  2.510$\pm$0.010 &  2.566$\pm$0.004 &  2.556$\pm$0.003 &  2.509$\pm$0.001 &  2.965 \\
    VY~Car & subarray     &  4.497$\pm$0.010 &  4.591$\pm$0.010 &  4.580$\pm$0.007 &  4.505$\pm$0.006 &  4.527$\pm$0.006 &  2.391 \\
    SW~Vel & subarray     &  4.891$\pm$0.010 &  4.908$\pm$0.010 &  4.919$\pm$0.006 &  4.887$\pm$0.003 &  5.064$\pm$0.003 &  4.302 \\
     T~Mon & subarray     &  3.309$\pm$0.010 &  3.469$\pm$0.010 &  3.428$\pm$0.006 &  3.320$\pm$0.005 &  3.422$\pm$0.001 &  2.978 \\
    AQ~Pup & subarray     &  5.014$\pm$0.010 &  4.982$\pm$0.010 &  5.009$\pm$0.007 &  4.993$\pm$0.006 &  4.889$\pm$0.004 &  4.171 \\
$\ell$~Car & full frame   &  1.020$\pm$0.028 &  1.020$\pm$0.049 &  1.003$\pm$0.048 &  0.927$\pm$0.033 &  0.720$\pm$0.001 &  0.690$\pm$0.053 \\
     U~Car & subarray     &  3.543$\pm$0.010 &  3.694$\pm$0.010 &  3.654$\pm$0.007 &  3.535$\pm$0.005 &  3.187$\pm$0.006 &  0.454 \\
    RS~Pup & subarray     &  3.324$\pm$0.010 &  3.336$\pm$0.010 &  3.350$\pm$0.006 &  3.331$\pm$0.005 &  3.316$\pm$0.001 &  $-$0.068 \\
  HD183864 & subarray     &  4.281$\pm$0.010 &  4.311$\pm$0.011 &  4.321$\pm$0.010 &  4.310$\pm$0.007 &  4.247$\pm$0.002 &  3.959 \\
  HD182296 & subarray     &  4.006$\pm$0.010 &  4.076$\pm$0.010 &  4.098$\pm$0.009 &  4.038$\pm$0.006 &  3.993$\pm$0.002 &  3.361 \\
$\gamma$~Phe & subarray     & -0.627$\pm$0.048 & -0.526$\pm$0.149 & -0.504$\pm$0.121 & -0.531$\pm$0.211 &  \nodata         &  \nodata         \\
$\psi$~And & full frame   &  2.415$\pm$0.076 &  2.507$\pm$0.084 &  2.462$\pm$0.037 &  2.389$\pm$0.034 &  2.332$\pm$0.001 &  2.307$\pm$0.157 \\
\enddata
\tablecomments{Magnitudes without errors are lower limits (flux upper
  limits)} 
\label{tab-phot}
\end{deluxetable}

\clearpage



\begin{deluxetable}{lcccccccc}
\tabletypesize{\footnotesize}
\rotate
\tablewidth{0pt}
\tablecaption{PL relation coefficients for different samples}
\tablehead{
  & \multicolumn{2}{c}{``New'' distances\tablenotemark{a}}
  & \multicolumn{2}{c}{Astr. distances\tablenotemark{b}}
  & \multicolumn{2}{c}{``Old'' distances\tablenotemark{c}} 
  & \multicolumn{2}{c}{Model (Z=0.02, Y=0.28)\tablenotemark{d}}\\
  \colhead{Band} &
  \colhead{$\alpha$} &
  \colhead{$\beta$} &
  \colhead{$\alpha$} &
  \colhead{$\beta$} &
 \colhead{$\alpha$} &
  \colhead{$\beta$} &
 \colhead{$\alpha$} &
 \colhead{$\beta$}}
\startdata
3.6 & $-3.47\pm0.06$ & $-5.72\pm0.07$
    & $-3.16\pm0.22$ & $-5.74\pm0.18$
    & $-3.54\pm0.04$ & $-5.76\pm0.04$
    & $-3.13\pm0.01$ & $-5.79\pm0.01$ \\
4.5 & $-3.38\pm0.06$ & $-5.68\pm0.07$
    & $-3.06\pm0.23$ & $-5.74\pm0.19$
    & $-3.45\pm0.04$ & $-5.72\pm0.04$
    & $-3.04\pm0.01$ & $-5.71\pm0.01$ \\
5.8 & $-3.44\pm0.06$ & $-5.68\pm0.07$
    & $-3.10\pm0.23$ & $-5.75\pm0.19$
    & $-3.51\pm0.03$ & $-5.70\pm0.04$
    & $-3.07\pm0.01$ & $-5.76\pm0.01$ \\
8.0 & $-3.46\pm0.06$ & $-5.73\pm0.07$
    & $-3.16\pm0.22$ & $-5.80\pm0.18$
    & $-3.57\pm0.03$ & $-5.74\pm0.04$
    & $-3.12\pm0.01$ & $-5.80\pm0.01$ \\
24  & $-3.52\pm0.06$ & $-5.71\pm0.06$
    & $-3.51\pm0.21$ & $-5.75\pm0.17$
    & $-3.67\pm0.03$ & $-5.78\pm0.03$
    & \nodata        & \nodata        \\
 70  & $-3.18\pm0.26$ & $-5.87\pm0.22$
    & $-3.34\pm0.27$ & $-5.87\pm0.23$
    & $-3.10\pm0.25$ & $-5.89\pm0.22$
    & \nodata        & \nodata        \\
\hline
$K$ & $-3.37\pm0.06$ & $-5.65\pm0.02$
    & $-3.32\pm0.12$ & $-5.71\pm0.03$
    & $-3.67\pm0.12$ & $-5.69\pm0.03$
    & \nodata        & \nodata        \\
\enddata
\tablenotetext{a}{Distances from \citet{fouque2007},
  \citet{benedict2007} or \citet{vanleeuwen2007}} 
\tablenotetext{b}{Distances from \citet{benedict2007} or
  \citet{vanleeuwen2007}} 
\tablenotetext{c}{Distances from \citet{storm2004},
  \citet{groenewegen2004}, \citet{barnes2005}, \citet{benedict2007},
  \citet{vanleeuwen2007} or this work (Table~\ref{tab-extrad})} 
\tablenotetext{d}{Based on the models described in
  Section~\ref{sec-discussion}} 
\label{tab-PL}
\end{deluxetable}

\clearpage





\clearpage

\clearpage

\clearpage

\clearpage

\clearpage

\clearpage

\clearpage

\clearpage

\end{document}